\documentclass[fleqn,usenatbib]{mnras}

\usepackage{amsmath} 
\usepackage{fourier}

\usepackage[english]{babel}															
\usepackage[protrusion=true,expansion=true]{microtype}	

\usepackage[pdftex]{graphicx}	
\usepackage{subcaption}
\usepackage{url}
\usepackage[thinc]{esdiff}
\usepackage{physics}
\usepackage[normalem]{ulem}

\usepackage[T1]{fontenc}

\DeclareRobustCommand{\VAN}[3]{#2}
\let\VANthebibliography\thebibliography
\def\thebibliography{\DeclareRobustCommand{\VAN}[3]{##3}\VANthebibliography}

\usepackage{xcolor}

\title[Ionisation of inner T Tauri star discs]{Ionisation of inner T Tauri star discs: effects of in-situ energetic particles produced by strong magnetic reconnection events}

\author[V. Brunn et al.]{
V. Brunn,$^{1}$\thanks{E-mail: Valentin.Brunn@umontpellier.fr}
A. Marcowith,$^{1}$
C. Sauty,$^{2,1}$
M. Padovani,$^{3}$
Ch. Rab$^{4,5}$
and C. Meskini$^{1}$
\\
$^{1}$Laboratoire Univers et Particules de Montpellier, Universit\'e de Montpellier/CNRS, place E. Bataillon, cc072, 34095 Montpellier, France\\
$^{2}$Laboratoire Univers et Th\'eories, Observatoire de Paris, Université PSL, Universit\'e Paris Cit\'e, CNRS, F-92190 Meudon, FranceÒ\\
$^{3}$INAF-Osservatorio Astrofisico di Arcetri, Largo E. Fermi 5, 50125 Firenze, Italy\\
$^{4}$Universit\"ats-Sternwarte, Fakult\"at f\"ur Physik,   Ludwig-Maximilians-Universit\"at M\"unchen, Scheinerstr.~1, 81679 M\"unchen, Germany\\
$^{5}$ Max-Planck-Institut für extraterrestrische Physik, Giessenbachstrasse 1, 85748 Garching, Germany\\
}

\date{Accepted XXX. Received YYY; in original form ZZZ}

\pubyear{2022}

\begin{document}
\label{firstpage}
\pagerange{\pageref{firstpage}--\pageref{lastpage}}
\maketitle

\begin{abstract}
Magnetic reconnection is one of the major particle acceleration processes in space and astrophysical plasmas. Low-energy supra-thermal particles emitted by magnetic reconnection are a source of ionisation for circumstellar discs, influencing their chemical, thermal and dynamical evolution. The aim of this work is to propose a first investigation to evaluate how energetic particles can propagate in the circumstellar disc of a T Tauri star and how they affect the ionisation rate of the disc plasma. To that end, we have collected experimental and theoretical cross sections for the production of H$^+$, H$_2^+$ and He$^+$ by electrons and protons. Starting from theoretical injection spectra of protons and electrons emitted during magnetic reconnection events, we have calculated the propagated spectra in the circumstellar disc considering the relevant energy loss processes. We have considered fluxes of energetic particles with different spectral indices and different disc magnetic configurations, generated at different positions from the star considering the physical properties of the flares as deduced from the observations obtained by the Chandra Orion Ultra Deep point source catalogue. We have then computed the ionisation rates for a disc whose structure has been calculated with the radiation thermo-chemical code {\tt ProDiMo}. We find that energetic particles are potentially a very strong source of local ionisation with ionisation rates exceeding by several orders of magnitude the contribution due to X-rays, stellar energetic particles and radioactivity in the inner disc.
\end{abstract}




\section{Introduction}

The way T Tauri stars accrete infalling surrounding matter through the transport of angular momentum is still a widely open subject in astrophysics. There are two main models that can explain the transport of angular momentum towards the outer disc, turbulent viscosity \citep{Shakura73, Shakura76} and magneto-centrifugal winds \citep{Blandford82,bai2016toward}. Ultimately both mechanisms involve an interaction between magnetic fields and ionised disc matter. Up to now, the best mechanism invoked to trigger the turbulence that all models need is the  magneto-rotational instability \citep{Balbus91}. The ionisation degree of the disc plays a crucial role, since it regulates the coupling between magnetic field and disc material (gas and dust) and it appears to be crucial to understand the accretion process around T Tauri stars \citep{Balbus03, sauty2019jet, Jacquemin-Ideetal19, Ray21, Jacquemin-Ideetal21}.

There are several sources of ionisation in the disc, each likely to dominate in different parts of it \citep{Cleeves13,Rab17}. X-ray and UV photons emitted by the star are quickly absorbed by the disc, thus significantly ionising only the disc surface. Conversely, radioactivity decay photons and cosmic rays (CRs) can ionise deeper disc layers. \citet{Umebayashi09,cleeves2013radionuclide} show that the ionisation rates per molecular hydrogen, called $\zeta$, due to radioactivity are, of the order of $(0.7-1)\times10^{-18}~{\rm s}^{-1}$. T Tauri stars as for the Sun are known to produce strong winds, which modulate the flux\footnote{Also know as spectrum, the flux is the number of particles per unit energy, time, area, and solid angle.} of incoming Galactic Cosmic Rays (GCRs). \citet{Cleeves13} investigate this effect using different scenarios. 
They first consider the case of a wind similar to our Sun and find an ionisation rate at 1 au between 0.8 and $3\times10^{-18}~{\rm s}^{-1}$ depending on the solar cycle. They further consider the possibility that T Tauri stars can induce stronger modulation. To evaluate this effect they use the stellar surface spot coverage as a proxy of the stellar activity. This assumption allows a calculation of the modulation potential in the force-free wind approximation \citep{2005JGRA..11012108U}. These models obtain ionisation rates below $10^{-20} ~ \rm s^{-1}$
This confirms that GCR modulation is strongly dependent on the properties of T Tauri stellar winds. For instance, \citet{Struminsky19} conclude that T Tauri stars can modulate the local cosmic ray flux beyond the outer radius of the disc at $\sim 10^{3}$ au and up to TeV energies. In contrast to this low ionisation rate trend, \citet{Padovani18} show that, accounting for the production of secondary particles and UV photons produced during the GCR propagation, higher ionisation rates can be maintained at higher levels compared to UV and X-rays up to large column densities of the order of or exceeding $10^{25} \rm{cm}^{-2}$. An alternative source of energetic supra-thermal charged particles can be associated to the high magnetic activity of T Tauri stars compared to the Sun. T Tauri stars are expected to produce fluxes of energetic particles about 5 orders of magnitude larger than the contemporary Sun \citep{Feigelson99}. \citet{Rodgers-Lee17} and \citet{Rab17} study the ionisation of the disc by the CRs emitted by the star itself. They find that ionisation rates by energetic particles ejected from the star are higher than ionisation rates by unmodulated GCRs up to a radius of $\sim$ 1 au. \citet{2018ApJ...853..112F} investigate the diffusive dilution of stellar energetic particles in the disc induced by the presence of a stochastic magnetic component sampled over a Kolmogorov spectrum, the background magnetic component being calculated using magnetohydrodynamic (MHD) simulations. The authors find high ionisation rates in the range $10^{-12}-10^{-9}~ \rm s^{-1}$ over a few tens of stellar radii displayed in patchy non-axisymmetrical structures. \citet{2020MNRAS.491.4742R} continue on their previous work by also considering a diffusive transport of stellar energetic particles into the disc. The ionisation rate due to stellar energetic particles dominate over GCR contribution up to large distances $\sim 70$ au. The authors confirm that these particles are a potential strong source of ionisation of the T Tauri star inner disc.

As previously stated, the accretion processes involved in circumstellar discs are strongly influenced by the ionisation rates. As GCRs are attenuated, taking into account ionisation by in-situ produced particles can have consequences for magnetohydrodynamic (hereafter, MHD) simulations. So far MHD simulations studying the star-disc interaction assume a totally ionised plasma \citep{matt2003collimation,ouyed2003three,kigure2005three,fendt2006collimation,orlando2011mass,zanni2009mhd,zanni2013mhd,ireland2020magnetic,ireland2022magnetic}. This assumption stems from the fact that close to the star the plasma is hot enough \citep{king2007accretion}. However, at larger radii where the accretion disc becomes sufficiently cold it is unlikely the plasma to be fully ionised.
In these simulations the circumstellar disc is considered geometrically thin and threaded by a magnetic field which drives the magnetorotational instability \citep{king2007accretion}. The magnetic field also has a structural role by collimating the central jets launched from a bipolar stellar magnetosphere. \cite{matt2003collimation} and \cite{fendt2006collimation} study the poloidal collimation using 2.5D simulations and find it to work best for a magnetic profile $\lvert \boldsymbol{B}_p \rvert \sim r^{-\mu}$ with $\mu \leq 1.3$ where $r$ is the distance from the star in cylindrical geometry.

Further MHD simulations study the angular momentum evolution, \citep{zanni2009mhd,zanni2013mhd,ireland2020magnetic,ireland2022magnetic}. The purpose of these works was to explain the constant angular velocity distribution over a few million years. The removal of the accreted angular momentum can be due either to the magnetic disc-star connection (disc locking mechanism) or by the outflows from the disc and the star (magnetic braking) combined with the disc viscosity. Such simulations are based on different magnetic configurations that will be used here in a simplified parametric form. 

Some simulations explore the effects of flares on the star-disc interaction. For instance, \citet{orlando2011mass} and \citet{colombo2019new} studied how large flares at the corotation radius of the disc can induce mass accretion episodes onto the star. The star in \citet{orlando2011mass} is a dipole with a strength $B\approx 1 ~\rm kG$ on the stellar surface. The authors describe the dynamical evolution of the magnetic field configuration and the disc properties occurring during the flare event. They find that the disc material eventually evaporates under the effect of thermal conduction. This produces an over pressure wave propagating down (or up) to the other disc side itself inducing a funnel flow allowing disc material to be accreted onto the star.

In this paper, we propose an alternative source of ionisation in continuity with the work of \citet{orlando2011mass,colombo2019new}. We study the ionisation due to energetic particles produced in the star-disc system associated with flaring episodes produced by magnetic reconnection. The magnetic reconnection at the origin of these flares is due to the interaction of the magnetic field of the star and the magnetic field of the disc. The particles emitted in this region propagate along the magnetic field lines in and above the disc. They can penetrate directly into the disc without necessarily to cross the dipolar field as is the case for particles emitted by the star \citep{Rab17}.

\section{Method}
\subsection{Disc Model}
\subsubsection{\textsc{ProDiMo}}
\label{sec:prodimo}
In order to evaluate the column density explored by the energetic particles we consider a model generated by the thermochemical radiation code \textsc{ProDiMo}\footnote{\url{https://www.astro.rug.nl/~prodimo}~\mbox{Version: 1.0 7e3ecc64}} \citep[PROtostellar DIsk MOdel, ][]{woitke2009radiation,kamp2010radiation,thi2011radiation,woitke2016consistent,Rab18}. By performing a wavelength-dependent radiative transfer calculation, \textsc{ProDiMo} calculates the gas and dust temperatures and the local radiation field. The temperature is determined by the balance of heating and cooling processes based on chemical abundances. The chemical network contains 235 different chemical species and 3143 chemical reactions \citep{kamp2017consistent,Rab17}.

\textsc{ProDiMo} computes the chemical abundances and calculates the thermo-chemical disc structure taking into account the radiative transfer of an external X-ray flux. 
We also use \textsc{ProDiMo} to calculate the effect of X-rays, describing the procedure in Sect. \ref{section:X-ray flare model}. The main physical parameters of the disc model (kept fixed) are listed in Table \ref{table:model parameters}.
 
\begin{figure*}
\begin{subfigure}{.5\textwidth}
  \centering
  \includegraphics[width=1\linewidth]{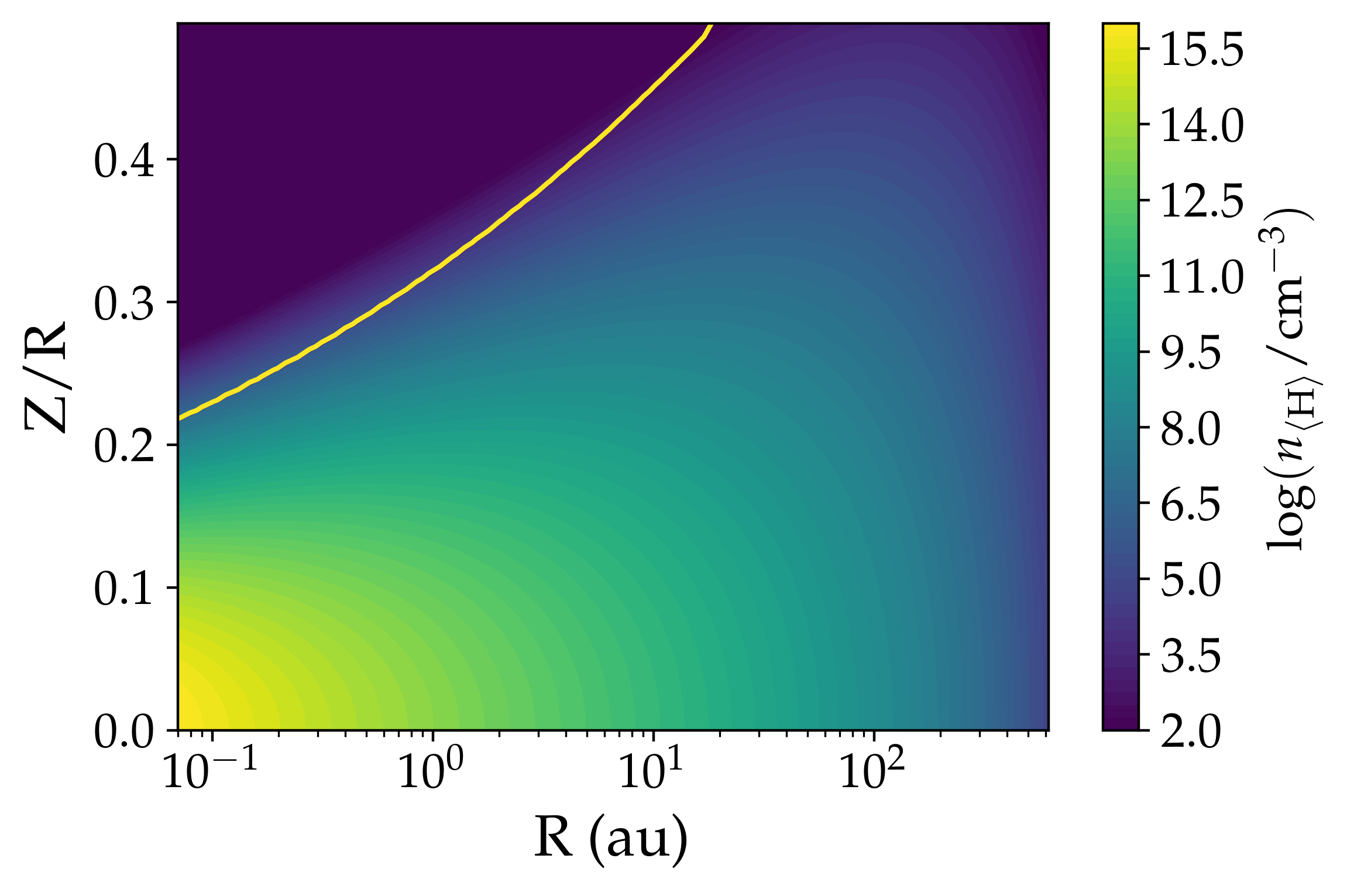}
  \caption{total hydrogen nuclei}
  \label{fig:gas distribution}
\end{subfigure}%
\begin{subfigure}{.5\textwidth}
  \centering
  \includegraphics[width=1\linewidth]{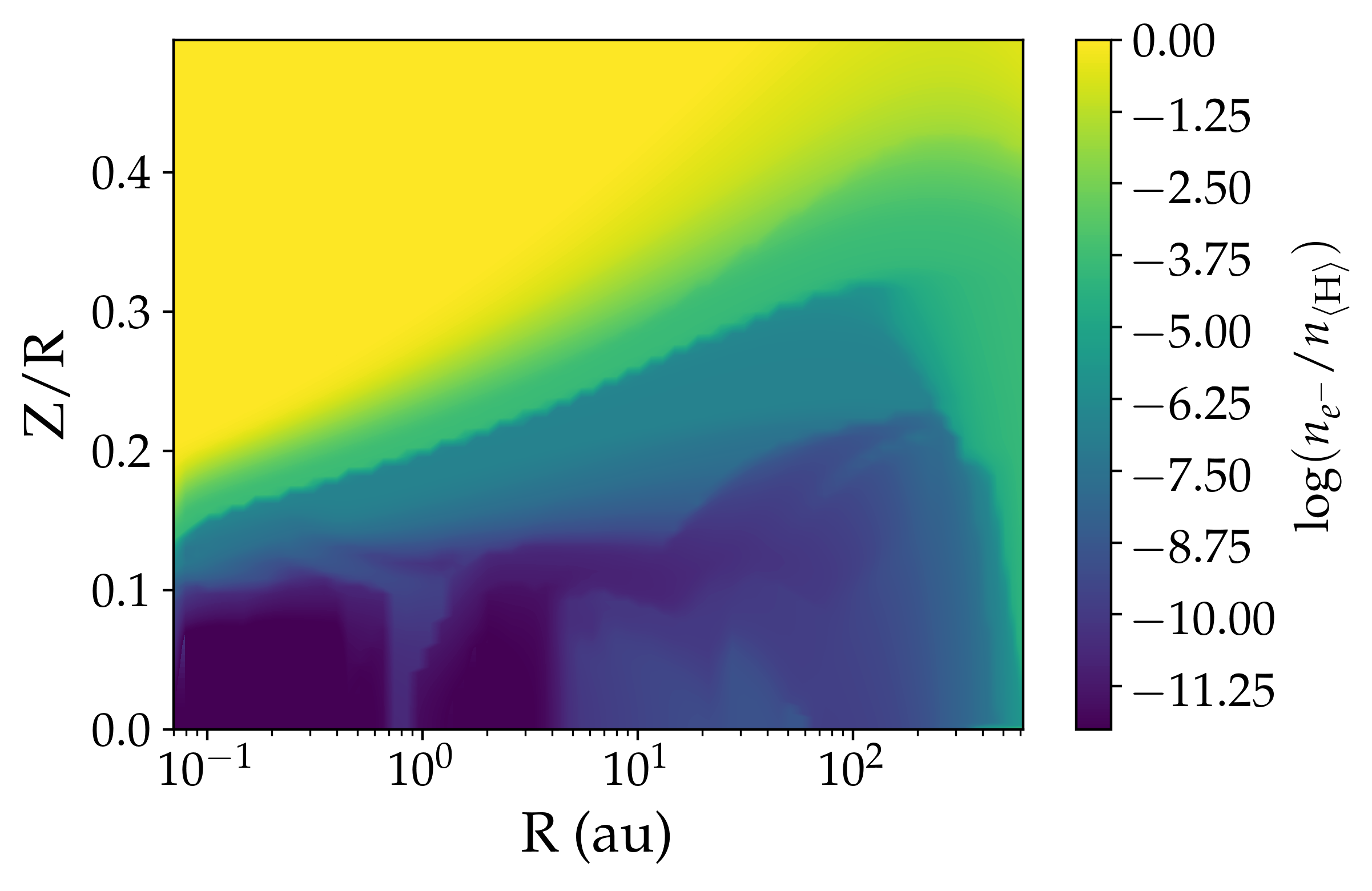}
  \caption{free electrons}
  \label{fig:electron distribution}
\end{subfigure}
\begin{subfigure}{.5\textwidth}
  \centering
  \includegraphics[width=1\linewidth]{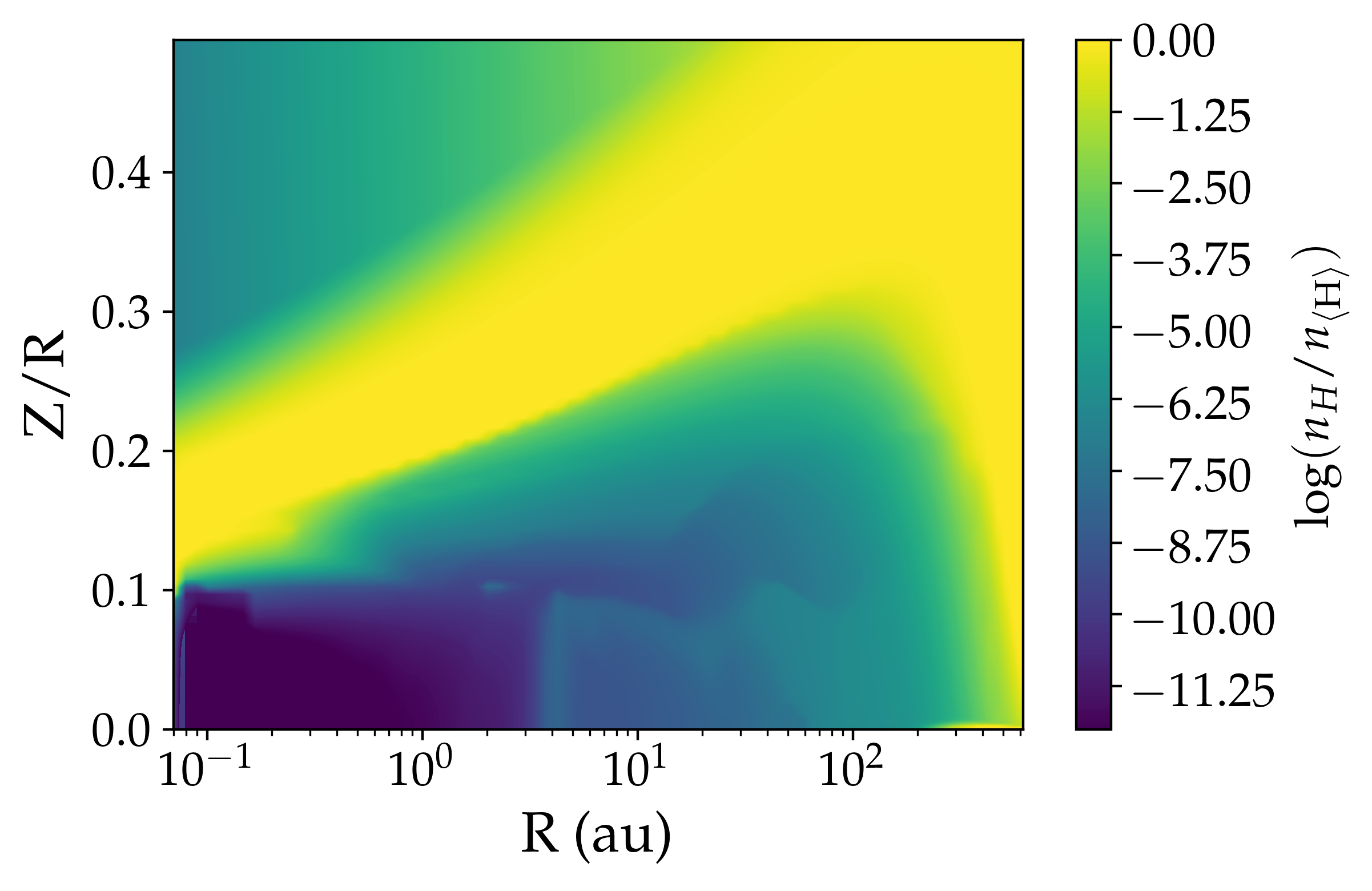}
  \caption{atomic hydrogen}
  \label{fig:H distribution}
\end{subfigure}%
\begin{subfigure}{.5\textwidth}
  \centering
  \includegraphics[width=1\linewidth]{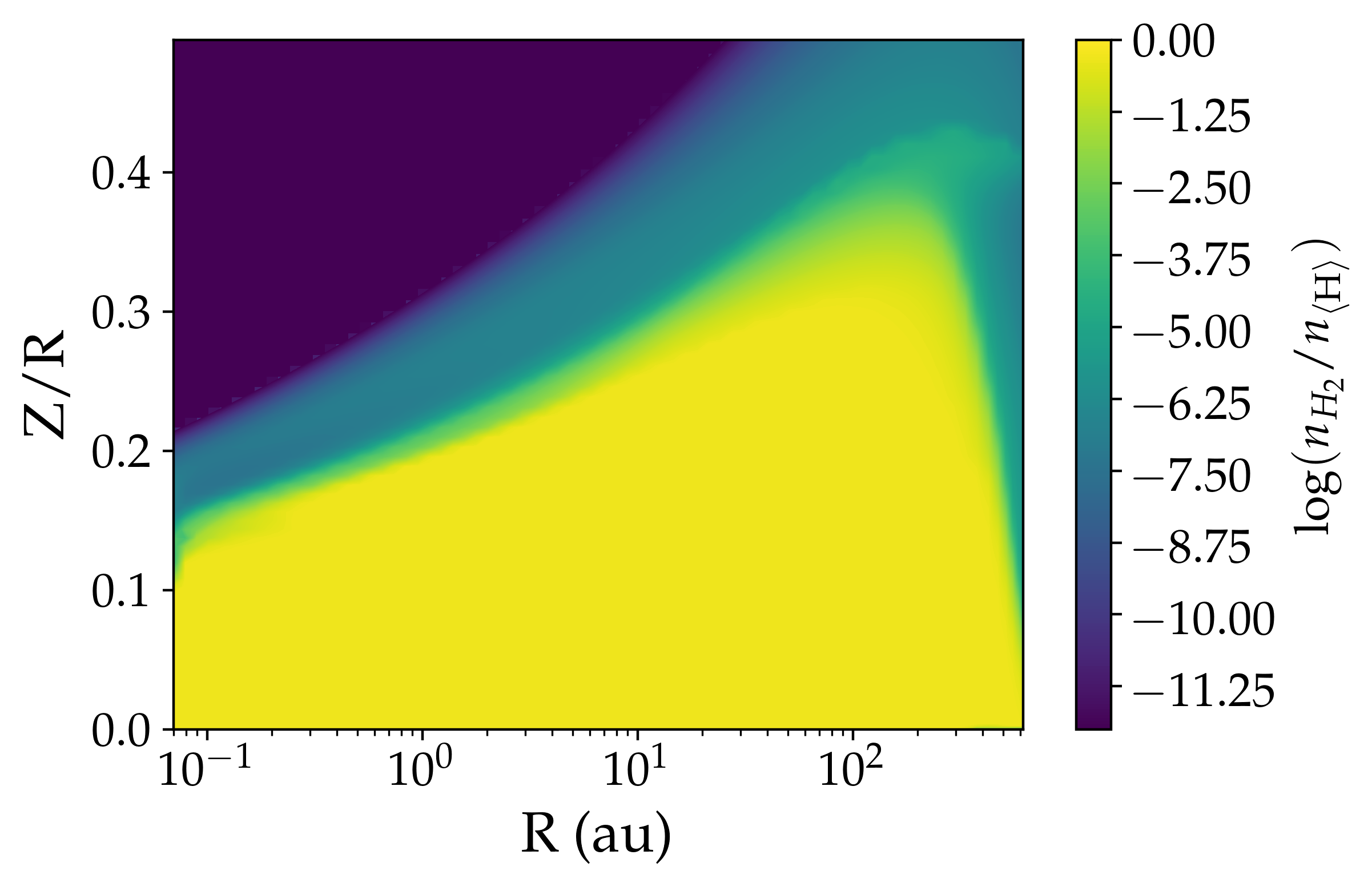}
  \caption{molecular hydrogen}
  \label{fig:H2 distribution}
\end{subfigure}
\caption{Two-dimensional structure of the accretion disc model. The height of the disc $Z$ is scaled to its radius ($Z/R$). The four panels show the density distribution of the gas particles (a), the free electrons (b), atomic hydrogen (c) and molecular hydrogen (d). The three last densities are scaled to the total gas number density. Along the yellow line in panel (a) the free electron density is equal to the atomic hydrogen density. We consider this line to be the boundary between the disc and its chromosphere, where we assume flares to occur.}
\label{fig:Number density distribution}
\end{figure*}

Figure~\ref{fig:Number density distribution} shows the number density of gas $n_{\rm \langle H \rangle}$ (total number of hydrogen nuclei), free electrons, atomic and molecular hydrogen for the reference model adopted in this work. This way of representing the distributions relative to the total gas density allows us to identify three distinct zones, each dominated by a chemical species. In each panel, the yellow zone delimits the part of the disc, where the fraction of the species represented is close to 1. 
The outer edges of the disc are composed of an electron-proton plasma. Then, moving deeper in the disc, electrons and protons recombine forming an atomic hydrogen layer. The inner part of the disc is denser and cooler and is essentially composed of molecular hydrogen.
To simplify our transport model, we consider the propagation of energetic particles in a disc composed of free electrons, hydrogen in atomic and molecular form, as well as atomic helium. \textsc{ProDiMo} provides the different distributions of the species of interest.

\begin{table}
\caption{Parameters of the disc model \citep{Rab17}. 
}
\label{table:diskmodel}
\centering
\begin{tabular}{l|c|c}
\hline\hline
Quantity & Symbol & Value  \\
\hline
stellar mass                          & $M_\mathrm{*}$                    & $0.7~\mathrm{M_{\sun}}$\\
stellar effective temp.               & $T_{\mathrm{*}}$                  & 4000~K\\
stellar luminosity                    & $L_{\mathrm{*}}$                  & $1.0~\mathrm{L_{\sun}}$\\
FUV excess                            & $L_{\mathrm{FUV}}/L_{\mathrm{*}}$ & 0.01\\
FUV power law index                   & $p_{\mathrm{UV}}$                 & 1.3\\
\hline

disc gas mass                         & $M_{\mathrm{disc}}$               & $0.01~\mathrm{M_{\sun}}$\\
dust/gas mass ratio                   & $d/g$                             & 0.01\\
inner disc radius                     & $R_{\mathrm{in}}$                 & 0.07~au\\
tapering-off radius                   & $R_{\mathrm{tap}}$                & 100~au\\
reference scale height                & $H(R=100\;\mathrm{au})$             & 10 au\\
                    
\hline

\end{tabular}

\label{table:model parameters}
\end{table}

\subsubsection{Magnetic field model}
\begin{figure}
    \centering
    \includegraphics[scale=0.6]{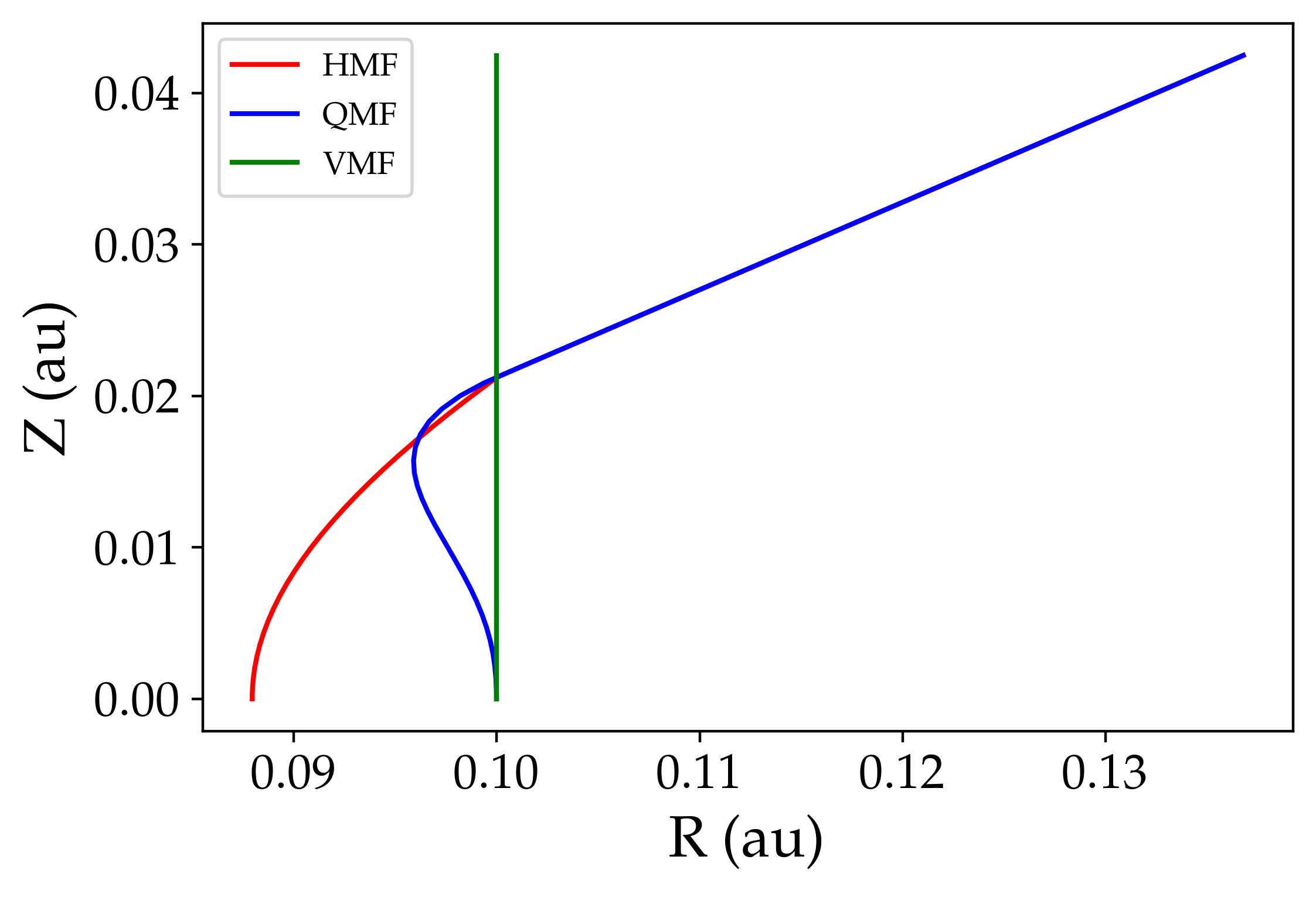}
    \caption{Example of the three poloidal magnetic field configurations as a function of the disc radius and height:
    vertical (VMF, solid green line), hyperbolic (HMF, solid red line), and quartic (QMF, solid blue line).
    }
    \label{fig: Magn conf}
\end{figure}

A limitation of the \textsc{ProDiMo} model is that it does not contain yet any magnetic field configuration. However, energetic particles are charged particles and the macroscopic structure of the magnetic field plays a major role in their propagation. Particles will follow the magnetic field lines and explore different regions of the disc depending on its magnetic configuration. The magnetic field pervading the accretion disc is expected to have a poloidal and a toroidal component \citep{Ferreira93}. For the poloidal component we consider three different configurations that are shown in Fig. \ref{fig: Magn conf}.

The simplest case considers the propagation along a vertical magnetic field line (VMF). This configuration will be our fiducial case hereafter. We further test the influence of more realistic magnetic field models. First we examine a hyperbolic magnetic field (HMF) configuration that makes an angle of 30° at the edge of the disc to allow magnetocentrifugal acceleration and the formation of winds, even if the disc is cold \citep{Blandford82}. Finally we considered a quartic magnetic field (QMF) configuration to account for a more efficient accretion on the outer layers of the disc and the possible formation of a dead zone in the equatorial plane \citep{Jacquemin-Ideetal21}.

\begin{figure}
    \centering
    \includegraphics[scale=0.6]{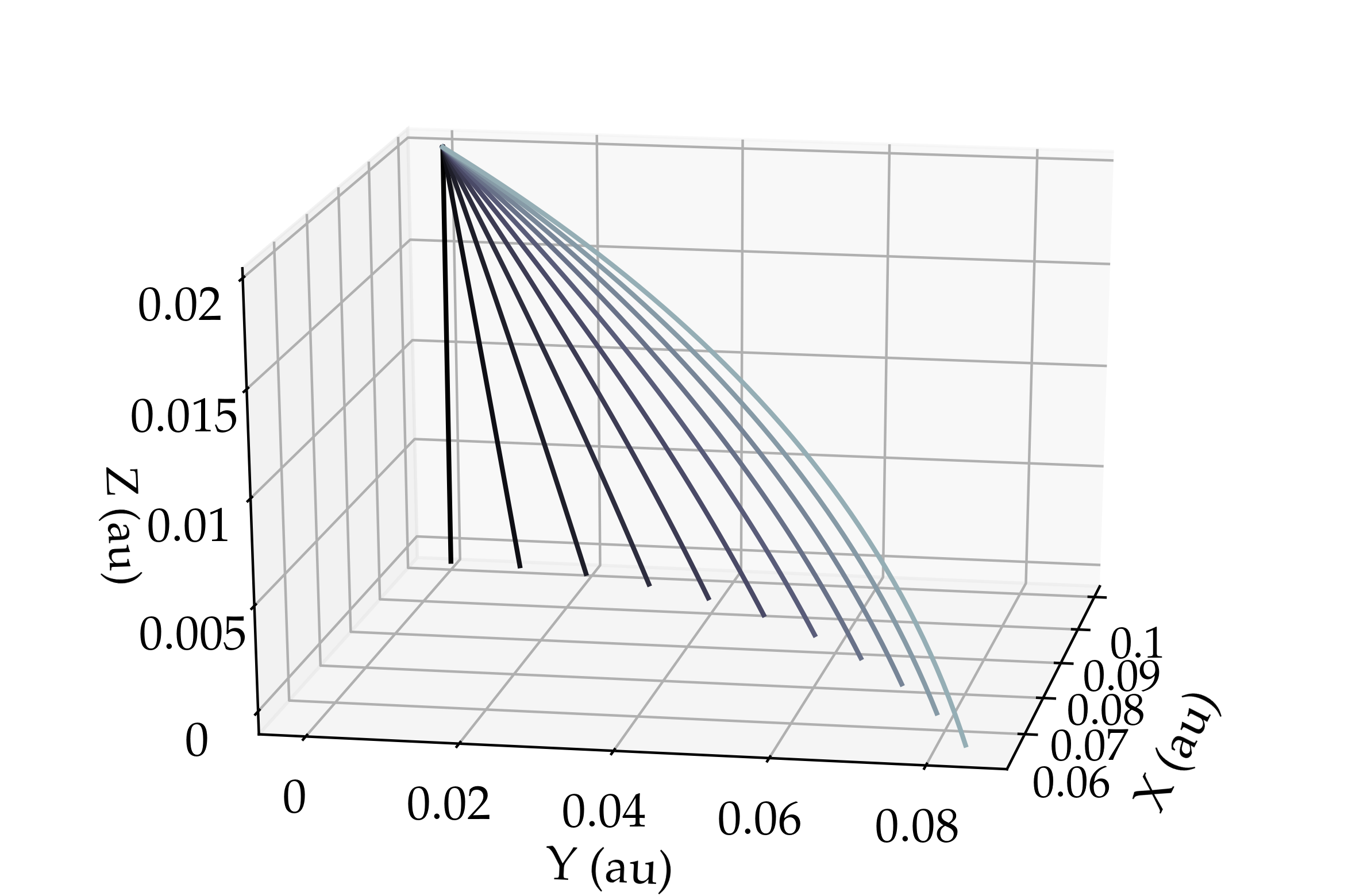}
    \caption{Illustration of the toroidal magnetic field configurations for the VMF configuration. The flare is located at the intersection of all lines. The parameter $b_g$ ranges from 0 (solid black line) to 1 (solid light grey line) in steps of 0.1.}
    \label{fig: Toroidal Magn conf}
\end{figure}

A toroidal component of the magnetic field is also included in our model. The strength of this component is parameterised by $b_g= B_{\rm \phi}/B_{\rm pol}$, the ratio of the toroidal to the poloidal component of the magnetic field. Magnetic fields with $b_g$ ranging from 0 to 1 are displayed in Fig. \ref{fig: Toroidal Magn conf}. This parameter is expected to have a substantial influence on the ionisation rate in the disc for two reasons. The first reason is that by propagating along different field lines, the particles explore different regions of the disc and different column densities, thus altering the propagated spectra and ultimately the ionisation rates. The second reason, which has the strongest impact on the ionisation rate, is the influence of the toroidal component of the magnetic field over the reconnection process itself. At the reconnection sites, the toroidal component of the magnetic field can approximately act as a guide field (see Sect. \ref{S:DIS}). Many studies have recently been carried out to investigate the influence of a guide field on the efficiency of the acceleration of non-thermal particles during a magnetic reconnection event, \citep{che2021ion,che2021formation,arnold2021electron,li2018large,stanier2017role}. It has been shown that the presence of a guide field suppresses the first-order Fermi acceleration process when $b_{\rm g} $ approaches 1, making the acceleration of non-thermal particles inefficient. Therefore, the injection flux of the particles is directly influenced by the intensity of the guide field.

Once the magnetic configuration is fixed, we superimpose it onto the disc model calculated by \textsc{ProDiMo}. 
Particles are injected at the position of the reconnection site and propagate along the magnetic field lines in the disc as test particles.

\subsection{Radiation and particle injection model}

Flares are explosive events that convert a fraction of the magnetic energy into particle kinetic energy through, in particular, magnetic reconnection. It is then expected that a significant fraction of the magnetic energy appear as a non-thermal population of energetic particles \citep{yamada2014conversion}. In this work, we only consider protons and electrons following a power-law distribution function. Flares also produce X-ray radiation, which is dominated by the bremsstrahlung radiation of hot thermal electrons. X-ray photons penetrate the disc first so their effects on the disc have to be taken into account before considering any effect due to energetic particles.

\subsubsection{X-ray flare model}
\label{section:X-ray flare model}
 We base our study on data provided by the Chandra satellite obtained in the Chandra Orion Ultradeep Project (COUP) sample, a sample of young stars in the Orion nebula \citep{getman2008a, getman2008b}. In order to account for the effect of X-rays we run \textsc{ProDiMo} including X-ray emission from a flare. The X-ray flux depends on several physical parameters such as the size of the reconnection site and the local plasma density. Using the  COUP data, we are able to correlate both emission size and gas density to the local gas temperature. We then derive an expression of the X-ray emissivity that depends only on the temperature. This emissivity is converted into a photon flux and used as an input for \textsc{ProDiMo} to derive a new disc model.

The thermal bremsstrahlung emissivity at frequency $\nu$ from an optically thin hot plasma emission is given in \citet{rybicki1991radiative} (their Eq. 5.14b). It reads,
\begin{equation}\label{Eq:ENU}
\epsilon _\nu =5.4\times 10^{-39} ~Z^2 n_e n_i T^{-1/2} e^{-h\nu / kT} \bar{g}_{\rm ff} ~\rm{erg~s^{-1}~Hz^{-1}~cm^{-3}~sr^{-1}} \ , 
\end{equation}
where $\bar{g}_{\rm ff}$ is the velocity averaged Gaunt factor, $n_e$, $n_i$ are respectively the local electron and ion densities, $Z$ is the effective charge of the ionised plasma, $h$ the Planck constant, $k$ the Boltzmann constant and $T$ is the temperature of the plasma. To compute $\bar{g}_{\rm ff}$, we use the parametric expressions derived in Eqs. (18) and (22) of \citet{van2014accurate}. To express the total power emitted by thermal bremsstrahlung we use Eq. (5.15b) in \citet{rybicki1991radiative}. This expression is obtained by integrating Eq. (\ref{Eq:ENU}) over the frequency and multiplying it by the volume of the flare $V_f$, 
\begin{equation}
L_X  = 1.4\times 10^{-27} Z^2 V_f n_e n_i T^{1/2} \bar{g}_{B}~\rm {erg ~s}^{-1} ,
\label{eq:Xluminosity}
\end{equation}
where $\bar{g}_{B}$ is the velocity averaged Gaunt factor averaged over frequencies, $\bar{g}_{B}$ is given by Eq.(26) of \citet{van2014accurate}.

The volume of the flare can be estimated as the cube of the typical length of the flare $V_f=L^3$. Assuming global neutrality $n_e=n_i$, the total X-ray luminosity is given in terms of the normalised values, 
\[
L_{10}=\frac{L}{10^{10} \rm cm} , \quad n_{e,10}=\frac{n_e}{10^{10} \rm cm^{-3}}, \quad T_{6}=\frac{T}{10^6 \rm K},   
\]
as
\begin{equation}
    L_X= 1.4 \times 10^{26} Z^2 L_{10}(T)^3 n_{e,10}(T)^2 T_{6}^{1/2} \bar{g}_{B}~\rm erg~s^{-1} \ .
\end{equation}

\begin{figure}
    \centering
    \includegraphics[scale=0.6]{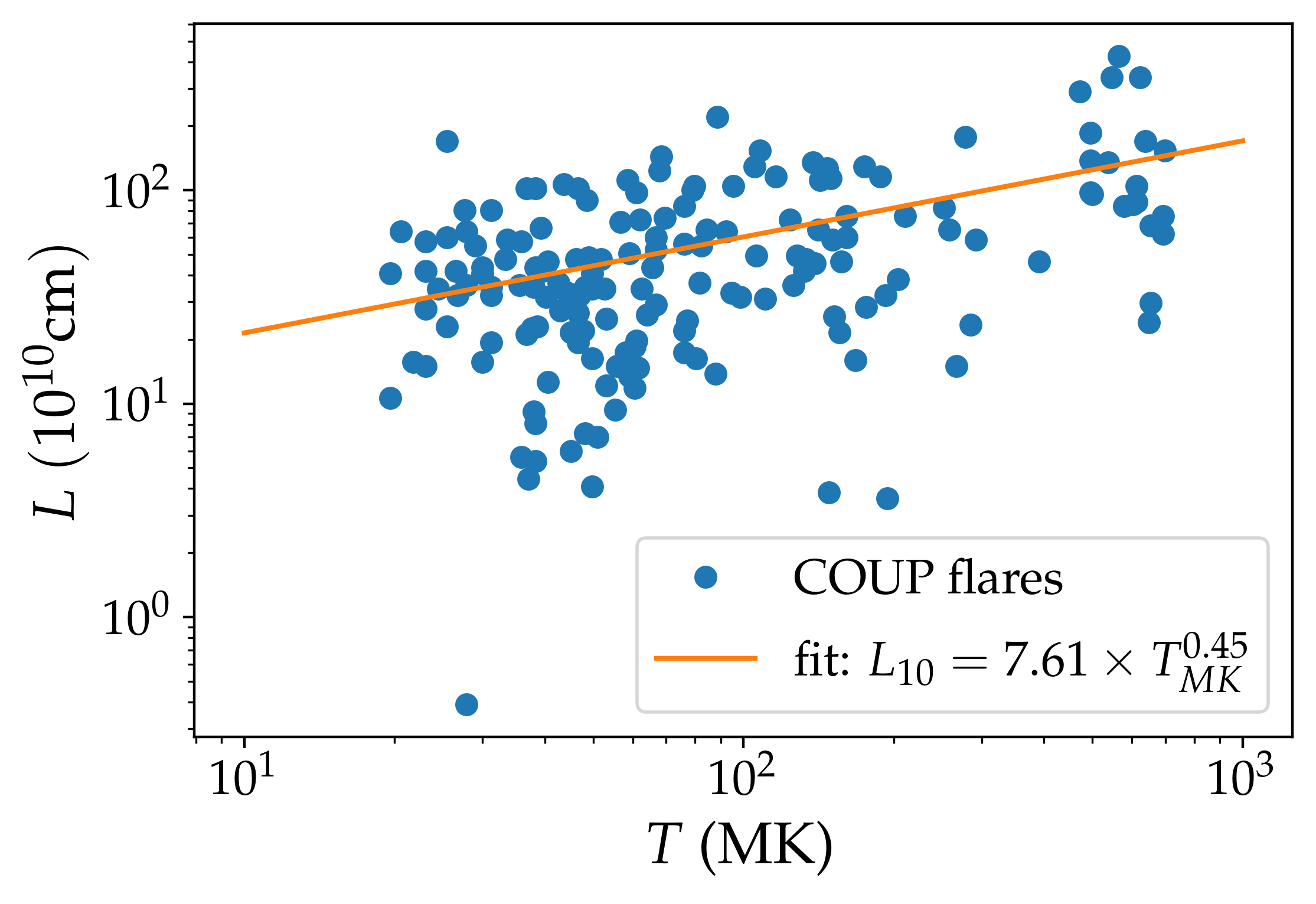}
    \caption{Flare length scale $L$ dependence as function of the flare peak temperature $T$ (orange fit) from the data derived by \citet{getman2008a} (blue dots).}
    \label{eq:L(T)}
\end{figure}

We derive $L_{10}(T)$ from the data of \citet{getman2008a}, see Fig. \ref{eq:L(T)}, which reads
\begin{equation}
\label{Eq:LT}
    L_{10}(T)=7.61 \times T_6^{0.45} \ .
\end{equation}

\begin{figure}
    \centering
    \includegraphics[scale=0.6]{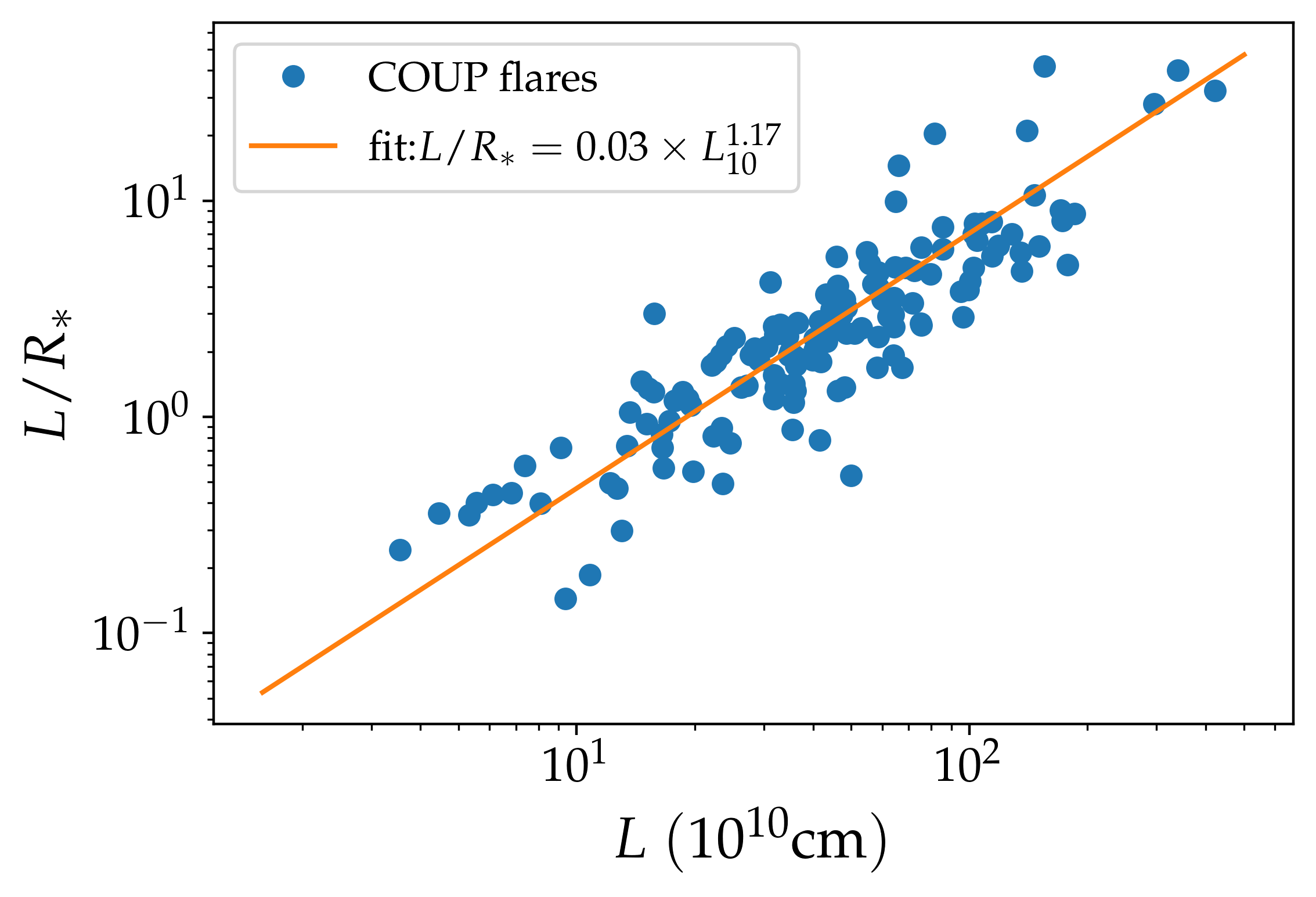}
    \caption{Comparison of inferred COUP flaring length to the length normalised to the stellar radius (orange line) using data from \citet{getman2008a} (blue dots).}
    \label{fig:LR(L)}
\end{figure}

The temperature is the observed X-ray temperature averaged over the entire flare loop. From the linear regression in Fig. (11) of \citet{getman2008b} the electron density $n_e$ can be expressed as a function of $(L/R_*)$ where $R_*$ is the stellar radius and
\begin{equation}
n_{e,10}(L/R_*)=4.60\times \left(\frac{L}{R_*}\right)^{-1.32} \ .
\end{equation}
Finally, from the data of \citet{getman2008a} (see Fig. \ref{fig:LR(L)}), an expression of $L/R_*$ can be found using a linear regression, 
\begin{equation}
     \frac{L}{R_*}=0.03 \times L_{10}^{1.17}.
     \label{eq:L/R(L)}
\end{equation}

Using the last three equations we can express the electron density in the flare area as a function of the temperature,
\begin{equation}
n_{e,10}(T)=20.5 \times T_{\rm 6}^{-0.69} .  
\label{eq:n(T)}
\end{equation}

The flare emissivity $\epsilon_{\nu}$ can be expressed with the temperature as the only parameter using Eq. \eqref{eq:n(T)}. This expression is derived using data from the COUP sample, so it applies only to this sample.

\begin{equation}
\epsilon _\nu =2.27\times 10^{-19} Z^2 \times T_{6}^{-1.88} e^{-h\nu / kT} \bar{g}_{\rm ff}~\rm{erg~s^{-1}~ cm^{-3}~ Hz^{-1}~sr^{-1}} \ .  
\label{eq:enu(T)}
\end{equation}

\begin{figure}
    \centering
    \includegraphics[scale=0.3]{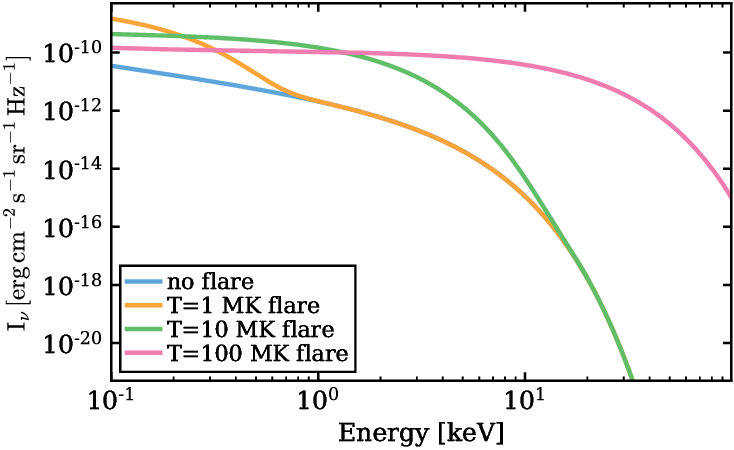}
    \caption{Specific intensity, $I_\nu = \epsilon_\nu L$, produced by 1, 10 and 100 MK flares (orange, green and magenta curves, respectively). These spectra are used as an input in \textsc{ProDiMo} to derive a new disc solution. The blue curve shows the case without flare.} 
    \label{fig:Xrayspactra}
\end{figure}

Figure \ref{fig:Xrayspactra} shows three different spectra corresponding to bremsstralhung emission at $1$, $10$, $100$ MK respectively.

The X-ray luminosity in Eq. \eqref{eq:Xluminosity} can be expressed as a function of the temperature only, 
\begin{equation}
    L_X=2.47\times 10^{31} \times T_{6}^{0.47} \bar{g}_{B} ~\rm{erg ~ s^{-1}} \ ,
    \label{eq:LX(T)}
\end{equation}
The flare luminosities corresponding to the temperatures studied are listed in Tab. \ref{tab:temperature-luminosity}.

Using Eqs. (\ref{eq:enu(T)}) and (\ref{eq:LX(T)}) we compute the density profile of the X-ray irradiated disc for the three flares shown in Fig. \ref{fig:Xrayspactra}. The disc structure that we obtain is determined by the X-ray emission of the flare before considering any energetic particle injection. However, we note that currently \textsc{ProDiMo} cannot calculate the effect of an X-ray source emitting from elsewhere than from the central star. Besides, as the flare we modelled is produced during the interaction between the magnetic fields of the disc and the star\footnote{we place it where the number density of atomic hydrogen and free electrons are equal, at the yellow line of Fig. \ref{fig:gas distribution}}, we underestimate the local energy deposition by the X-ray photons. Our conclusions could be slightly sensitive to this effect. We further discuss this aspect in Sect. \ref{ss:Radiation model}. An updated version of the \textsc{ProDiMo} code is necessary to account for this more refined X-ray radiative transfer effect and will be considered in a future work. 

\begin{center}
\begin{table}
\caption{Density, luminosity and position of flares of different temperature.}
\label{tab:temperature-luminosity}
\begin{tabular}{ |l| c| c| c |}
    \hline
   Flare temperature (MK) & $1$ & $10$ & $100 $ \\ \hline
   Luminosity (erg s$^{-1}$) & $3.55\times 10^{31}$ & $9.87\times 10^{31} $& $2.67\times 10^{32} $\\ \hline
\end{tabular}
\end{table}
\end{center}

\subsubsection{Energetic particle injection model} \label{S:INJ} 
For both electrons and protons, we express the energy distribution per unit volume $F(E)$ (hereafter energy distribution) or the flux $j(E)$, in terms of the particle kinetic energy $E$,
where $j(E)= F(E) \frac{v}{ 4\pi}$, $v$ is the particle speed. The flux $j$ has units ($ \rm particles ~ s^{-1} cm^{-2} sr^{-1} eV^{-1}$) and is composed of a thermal and a non-thermal component. 

The thermal component is defined by its temperature $T$ and its normalisation $n_{\rm th}$. If we take equipartition between protons and electrons temperatures, considering  energetic electrons (k = e) and protons (k = p), we get,
\begin{equation}
\label{jTH}
  j_{k,\rm {th}}(E) =  n_{k,\rm th} {2 \over \sqrt{\pi}} \sqrt{{E \over k T}} \exp(-E/kT) {\beta_{k}(E) c \over 4\pi k T} \ .
\end{equation}
The peak value is reached at $E = 3/2 k T= E_{\rm th}$. 

As it is usually assumed in reconnection models for solar flares  \citep{ripperda2017reconnection} and supported by observation of solar flares \citep{emslie2012global,matthews2021high} energy equipartition between non-thermal protons and electrons below 1 GeV is a good proxy of non-thermal content energetics. The non-thermal particle energy density is,
\begin{equation}
U_{e,\rm{nt}}= U_{p,\rm{nt}}=\int_{E_{\rm c}}^\infty E F_p(E) \dd E=\int_{E_{\rm c}}^\infty E F_e(E) \dd E
\ ,    
\end{equation}
where the subscript 'nt' stands for non-thermal.
Assuming they have the same injection energy $E_{\rm c}$, electrons and protons have the same non-thermal energy distribution $F_e(E)=F_p(E)=F(E)$, so 
\begin{equation}
 j_{p}(E)= \frac{\beta_p(E)}{\beta_e(E)} j_{e}(E) 
\ .  
\label{eq:flux proton flux electron}
\end{equation}
Beyond energy equipartition, another possible assumption is a momentum equipartition between non-thermal particle species. In that specific case we find $j_p(E)\ll j_e(E)$ for $E< 1 ~\rm GeV$ so the ionisation rates are governed by the electron flux. Both assumption lead essentially to the same results.

The non-thermal flux is defined by its normalisation $j_{\rm nt,0}$, which is fixed by the injection energy $E_{\rm c}$. The non-thermal component results from the magnetic reconnection process so we can introduce two other energies, a break energy $E_{\rm b} > E_{\rm c}$ and a maximum energy $E_U$. The distribution is a power-law with indices $\delta_1$ between $E_{\rm c}$ and $E_{\rm b}$ and $\delta_2$ between $E_{\rm b}$ and $E_{\rm U}$. This model is consistent with the non-thermal electron distribution observed in solar flares \citep{mewaldt2005proton}.

The injection energy is proportional to the thermal energy, $E_{\rm c} = \theta E_{\rm th}$. However, most of the models associated with T Tauri flares only impose a single power law. Thus, unless specified, we only consider $E_c$, $E_U$, $\delta_1$ and an exponential cut-off beyond $E_{\rm U}$,
\begin{equation}
\label{jNTH}
    j_{k,\rm nt}(E)=n_{k,\rm nt} \left(\frac{E}{E_{\rm c}}\right)^{-\delta}\exp\left(-\frac{E}{E_{\rm U}}\right) \frac{\beta_k(E)c}{4\pi }
    \,,
\end{equation}
where we have set $\delta=\delta_1$. The factor $\beta(E) = v(E)/c =\sqrt{((1+\bar{E})^2-1)/(1+\bar{E})^2}$ with $\bar{E}=E/mc^2$. The effect of an intermediary spectral break will be discussed in section \ref{S:DIS}.

The density in Eq. (\ref{eq:n(T)}) is the sum of the thermal and the non-thermal components, $n_{k}= n_{k,\rm th}+ n_{k,\rm nt}$. Using the equality $j_{\rm nt}(E_{\rm c})= j_{\rm th}(E_{\rm c})$, we have a way to derive $n_{k,\rm nt}$. 
In order to do so, we equal the non-thermal and the thermal fluxes, Eqs. (\ref{jTH}) and \eqref{jNTH}, respectively, to the injection energy, $E=E_c=\theta E_{\rm th}$,
\begin{equation}
    n_{k,\rm nt}=n_{k,\rm th} {2 \over \sqrt{\pi}} \frac{1}{k_{\rm B} T}\sqrt{3/2\theta} \exp{(-3/2\theta)}.
    \label{eq:nNth}
\end{equation}
Since $n_k=n_{k,\rm nt}+n_{k,\rm th}$,
\begin{equation}
    n_k=n_{k,\rm th}\left(1+{2 \over \sqrt{\pi}} \sqrt{3/2\theta}\frac{1}{k_{B\rm} T} \exp{(-3/2\theta)}\right)\ ,
\end{equation}
it follows using Eq.\eqref{eq:n(T)} that,
\begin{equation}
    n_{k,\rm th}=\frac{20.5 \times T_{\rm 6}^{-0.69}}{1+{2 \over \sqrt{\pi}} \sqrt{3/2\theta}\frac{1}{k_{\rm B} T} \exp{(-3/2\theta)}} .
    \label{eq:nth(T)}
\end{equation}

The power-law index of the energetic particles accelerated in a magnetic reconnection event can take a wide variety of values from 1.5 to 9 \citep{oka2018electron}. This spread is first due to the different types of magnetic reconnection i.e collisional or non-collisional, see \citet{ji2011phase} for a review. The dispersion in power-law indices is also due to the variety of astrophysical plasmas and their properties i.e the $\beta_{p}$ parameter defined as the ratio of the gas pressure to the magnetic pressure and the plasma magnetisation $\sigma_{p}$ defined as the ratio of the magnetic energy to the enthalpy density of the plasma \citep{ball2018electron},
\[
    \beta_{p}=\frac{8\pi n k_{\rm B} T}{B_0^2}\qquad \text{and} \qquad \sigma_{p} = \frac{B_0^2}{4\pi n m_{p} c^2} \ ,
\]
where $m_p$ is the proton mass and $B_0$ is the background magnetic field amplitude. The value of the index also strongly depends on the magnitude of the guide magnetic field in numerical experiments.
The guide field is the magnetic field component perpendicular to the reconnection plane in 2D reconnection processes. In an axisymmetric configuration, this component is interpreted as the azimuthal component, $B_\phi$. If there is some small scale turbulence, the macroscopic mean azimuthal field gives an upper limit of the guide field magnitude.

In order to identify the type of magnetic reconnection that is expected to take place above the circumstellar disc of a T Tauri star, we use the phase diagram in \citet{daughton2012emerging}. The phase diagram determines whether the reconnection is non-collisional, collisional MHD with plasmoids \footnote{plasmoids are islands of magnetic field lines formed and ejected during magnetic reconnection, see \citet{loureiro2015magnetic} for a review.} or collisional in the Sweet-Parker regime. Different types of reconnection have different acceleration mechanisms, so it is important to estimate, which reconnection is expected to take place above the circumstellar disc of a T Tauri star.
There are two main controlling parameters in the phase diagram. The first parameter is the system size $L_0$. The second parameter is the global Lundquist number $S_\nu$ defined as,
\begin{equation}
    S_\nu=\frac{v_{\rm A} L_0}{\nu}, 
\end{equation}
where $v_{\rm A}$ is the local Alfv\'{e}n speed and $\nu$ the plasma resistivity. 

The above parameters can be expressed in terms of the temperature, the particle number density and the magnetic field strength at the flare location. Considering a typical flare size $L=10^{10} {\rm cm}= L_0$ \citep{getman2008a}, and typical values of density and temperature of the background plasma at the location of the flare, $n_{e}=10^{10} \rm cm^{-3}$ and $T = 1000$ K respectively, we find that magnetic reconnection occurs in the collisional regime (with plasmoids for a 2D geometry).

 \citet{arnold2021electron} explore the spectral index $\delta$ of the accelerated electrons in the collisional regime using 2D simulations that take into account the feedback of the accelerated particles. In the collisional regime, with an Alfv\'{e}n velocity $v_A=c/60 $ corresponding to a magnetisation $\sigma=2.8\times 10^{-4}$ and for a guide field $b_g=B_g/B_0$ between $0$ and $1$, their simulation indicates that $\delta$ lies between $2.5$ and $8$. Strong guide fields produce softer non-thermal spectrum. 
 
\begin{figure}
    \centering
    \includegraphics[width=1\linewidth]{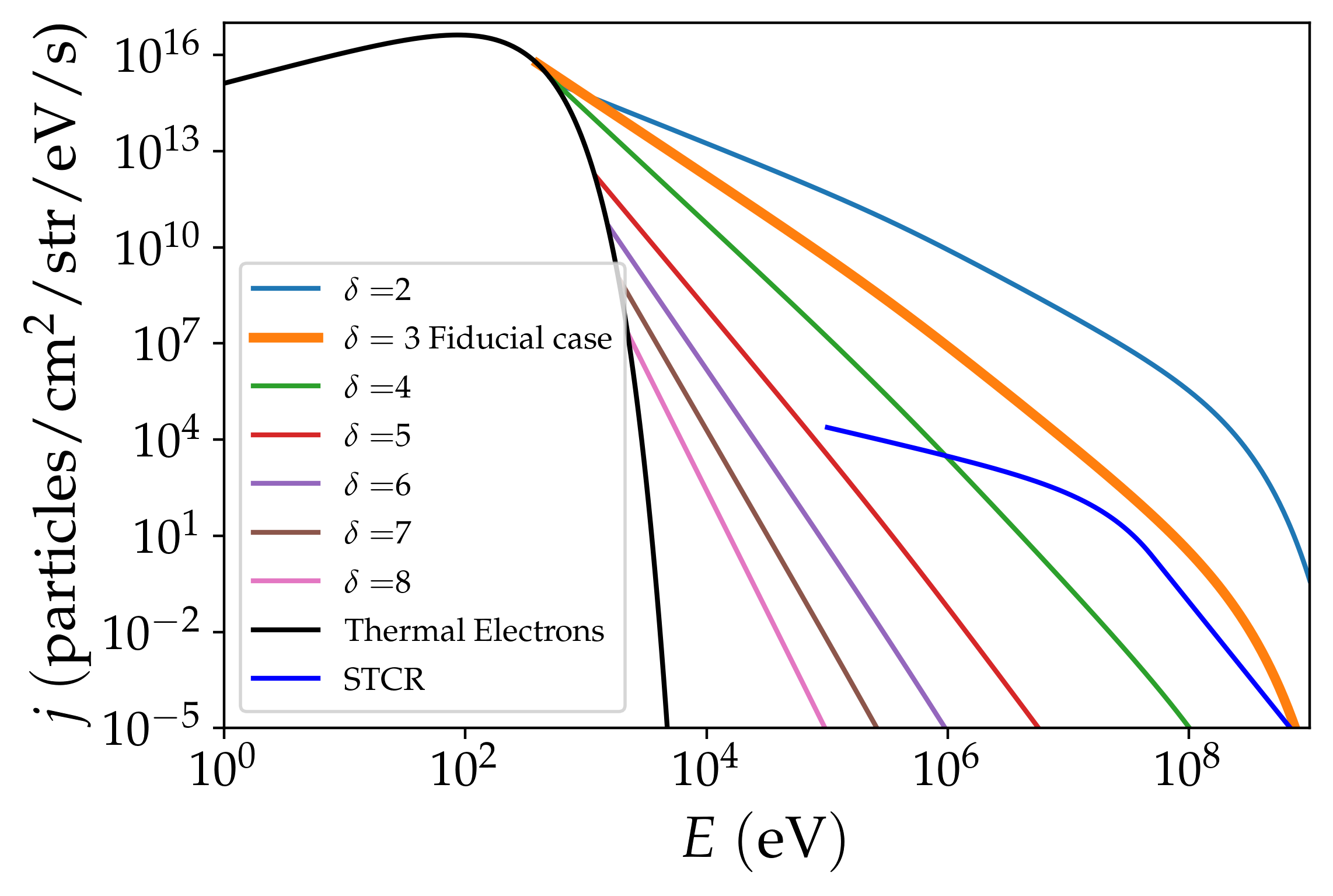}
    \caption{Electron injection spectra as function of their index in the case of a 1 MK Flare in the case of no guiding field. Proton spectra can be deduced from Eq. \ref{eq:flux proton flux electron}. For comparison, the proton flux (STCR) at $R=0.1 ~\rm au$ emitted by the stellar flare studied by \citet{Rab17} is plotted in deep blue.} 
    \label{fig:injection spectra}
\end{figure}

  \citet{mewaldt2005proton} measurements of solar flares from October 2003 indicate softer indices, rather approximately equal to $4$. \citet{waterfall2020predicting} conducting a study on non-thermal particle injection in young stellar object use $3$. Hereafter, we assume a spectral index $\delta = 3$ for the non-thermal component in our model, but softer and harder spectra effects will be discussed in the parametric study conducted in Sect.\ref{sec: Parametric study}. Taking into account the results of \citet{arnold2021electron} and also from the large uncertainty in the power-law index obtained by this mechanism we decide to narrow the range of $\delta$ from $1.5-9$ of \citet{oka2018electron} to $2-8$. The upper limit has been chosen in order to have a non-thermal component that substantially emerges from the thermal one (see the pink curve in Fig. \ref{fig:injection spectra}).

In Fig. \ref{fig:injection spectra} we show the different non-thermal electron spectra that are injected into the disc. To construct them we proceed in two steps. First we impose that the total flux of each non-thermal component is the same. To ensure this, the injection energy of the non-thermal particles is adjusted such that for each index $\delta$, we have,
\begin{equation}
    \int_{E_{\rm c}}^{\infty} j(E) \dd E = J_{\rm nt} \ .
    \label{eq:constant energy density}
\end{equation}
Second, to be considered as a part of the injection flux the particles of the non-thermal component must be dominant over the thermal component. 
The reference value for $J_{\rm nt}$ is the integrated flux injected by a flare with a spectral index $\delta=3$ and an injection energy of the non-thermal particles $E_{c}=3 E_{\rm th}$. Notice from Fig.\ref{fig:injection spectra} that typical injection energies are in the range 1-10 keV, these are consistent with low energy cut-offs of the non-thermal particle distribution deduced from solar flare surveys, e.g. \citep{2016ApJ...832...27A}. We have considered the same injection energy and the same maximal energy for electrons and protons and we discard any break energy unless otherwise specified.
By default at high energy we fix $ E_{U}= 100$ MeV and we add an exponential cut-off energy $\exp (-E/E_U)$. This will be our fiducial injection flux hereafter. The fiducial non-thermal flux is therefore
\begin{equation}
    j_{k,\rm nth}=n_{k,\rm nt} {\beta_k(E) c \over 4\pi} \left(\frac{E}{3E_{\rm th}}\right)^{-3} \exp\left(-E/100~\rm{MeV}\right),
    \label{eq:injection spectrum}
\end{equation}
corresponding to the orange thick curve in Fig. \ref{fig:injection spectra}. The normalisation factor $n_{\rm nt}$ is given by Eqs. \eqref{eq:nNth} and \eqref{eq:nth(T)} with $\theta=3$. In the fiducial case the flare is located at 0.1 au from the star, has a temperature of 1 MK and the magnetic field is purely vertical and poloidal. We study the particle propagation in the disc in the next section.

\subsection{Propagation model}
In order to calculate the ionisation rate in the disc due to the energetic particles produced by the flare, it is necessary to study the propagation and attenuation by the disc material of the primary and secondary particle spectra. We define $j_{k}(E,N)$ as the flux of electrons and protons that have penetrated a column density $N$ after injection and we label the injection flux at the disc surface ($N=0$) as $j_{\rm e/p}(E,0)$, computed by Eq. \eqref{eq:injection spectrum}. 

To derive the propagated spectra, we use the continuous slowing-down approximation (CSDA) as in \citet{Padovani2009,Padovani18}.
Thus, the flux at a column density $N$ can be expressed as a function of the injection flux and the energy loss function (Sect. \ref{sec:energy loss functions}).

\subsubsection{Energy Loss functions}
\label{sec:energy loss functions}
The energy loss function $L(E)$ is constructed to follow the energy evolution of the particles entering in the disc. $L(E)$ is defined as the energy lost per unit column density,
\begin{equation}
\diff{E}{N}=-L(E) \ ,
\label{eq:def L}
\end{equation}
which depends both on the projectile (proton or electron) and on the target medium (H, H$_2$, He).
In our model, the electron loss function is deduced from the data provided by \citet{dalgarno1999electron} at low energy ($\rm <1keV$), which are based on theoretical calculations. At high energy ($ \rm >1keV$) we use the National Institute of Standards and Technology database \footnote{\url{https://physics.nist.gov/PhysRefData/Star/Text/ESTAR.html}\label{NIST}}. See \citet{padovani2022cosmic} for an updated description of the electron energy loss function. At low energy, energy losses in molecular gases are controlled by rotational and vibrational excitations, at intermediate energy by ionisation and at high energy by radiative losses, see Fig. \ref{fig:electron loss function}. The data of the Stopping and Range of Ion in Matter \citep{ziegler2010srim} are used to construct the proton loss functions. At high energies, above the threshold $E^\pi = 280 \rm MeV$, we add
energy losses due to pion production, as reported by \citet{Padovani18}, Fig. \ref{fig:proton loss function}. At very low energy ($<1-10 \rm eV$) Coulomb losses are dominant, we use the expression from \citet{schlickeiser2013cosmic}.

Another source of losses to be considered is the losses generated by the return current produced by the electrons, in case the proton component is negligible.
The return current is driven by an electric field induced in the plasma by the energetic electrons in the beam. This electric field, in addition to drive the return current, also extracts energy from the beam electrons. To estimate the importance of the return current losses we follow the procedure detailed in \citet{holman12}. 
We derive a characteristic column density from the electron injection site, $N_{rc}$ below which this loss mechanism is important. For electrons of energy less than 1 GeV, we find $N_{rc} \simeq 10^{19}~\rm  cm^{-2}$. Since we consider energetic particle injection at the edge of the disc minimum column densities of about $10^{19}~\rm  cm^{-2}$ are reached very quickly. If the flare would occur further away above the disc then the effect of return current may have to be considered properly. Its detailed modelling is postponed to a future work.

Since the abundances of H, H$_2$ and He vary as a function of the disc depth, we compute an average loss function dependent on the coordinate $s$ along a magnetic field line defined as

\begin{equation}
{\bar{L}(E,s) ={1 \over s} \sum_i \int_0^s L_i \dv{N_i(s)}{N(s)} \dd s} \ ,
\end{equation}
where $ i={\rm H, H_2, He}$ and $ \dd N_i(s)$ is the variation of column density of the species $i$ for a variation $\dd s$ and $\dd N(s)= \sum_i \dd N_i(s)$. Since $L_i$ is independent of position and $\dv{N_i}{N}$ can be expressed in terms of the number density as $\dv{N_i}{N}=\frac{n_i(s)}{n(s)}$, the ratio of densities of species $i$ over the total gas density. The average loss function $ \bar{L}$ can be rewritten as
\begin{equation}
 \bar{L}(E,s)=\sum_i f_i(s) L_i(E),
\end{equation}
\begin{figure}
    \centering
    \includegraphics[width=1\linewidth]{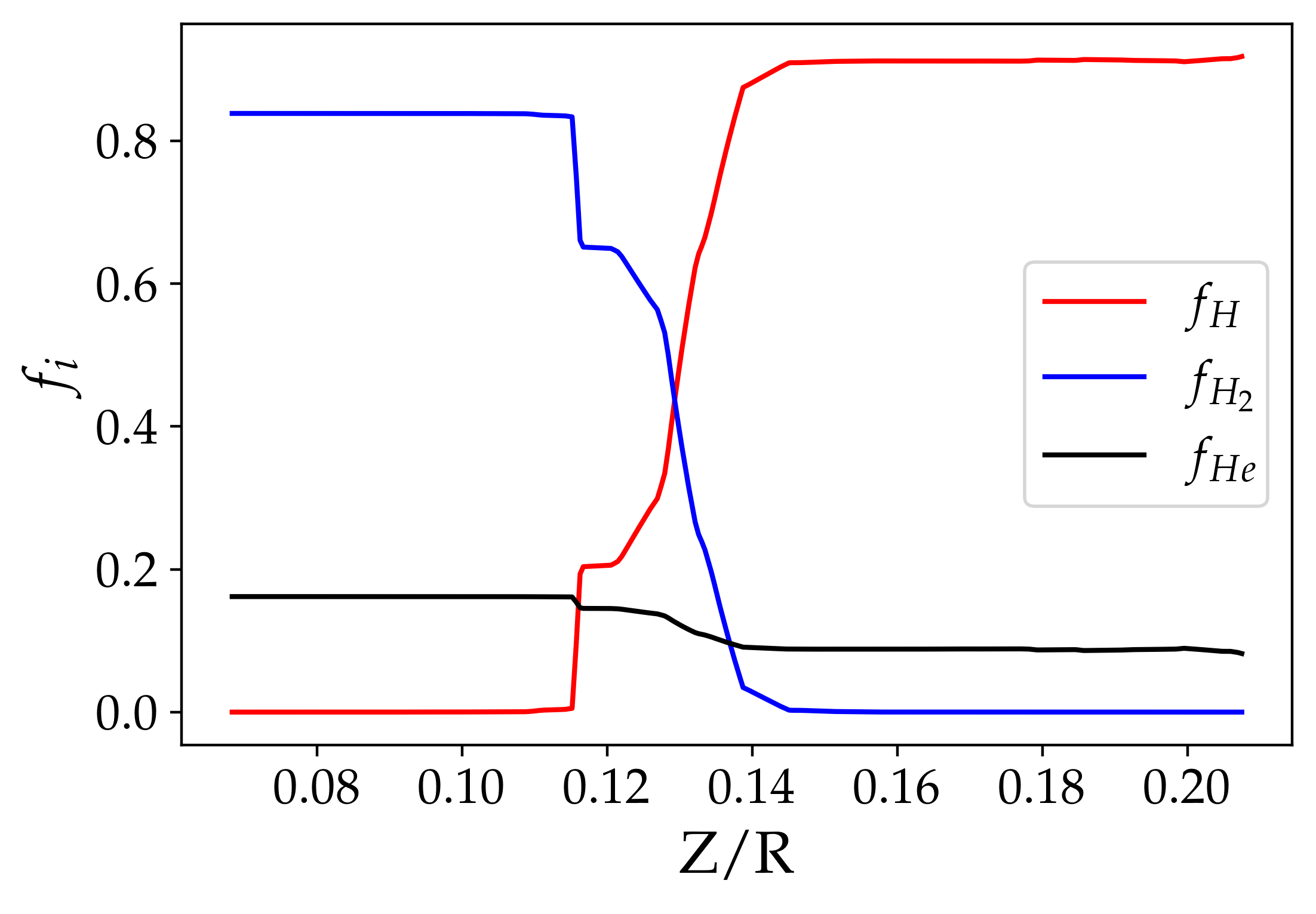}
    \caption{Plots of the fraction of chemical element weighted along the path as function of the depth in the disc, computed from equation \ref{Eq:Weight}. In red we plot the fraction of atomic hydrogen, in blue the fraction of molecular hydrogen and in black the fraction of helium.}
    \label{fig:f_i}
\end{figure}
where $f_i(s)$ is the fraction of species $i$ weighted along the path, given by
\begin{equation}\label{Eq:Weight}
f_i=\frac{1}{s}\int_0^s \frac{n_i(s')}{n(s')}\dd s',
\end{equation}
Fig. \ref{fig:f_i} shows the species fractions and Fig. \ref{fig:loss functions} the average loss functions for a set of column densities.
\begin{figure}
\begin{subfigure}{.5\textwidth}
  \centering
    \includegraphics[width=1\linewidth]{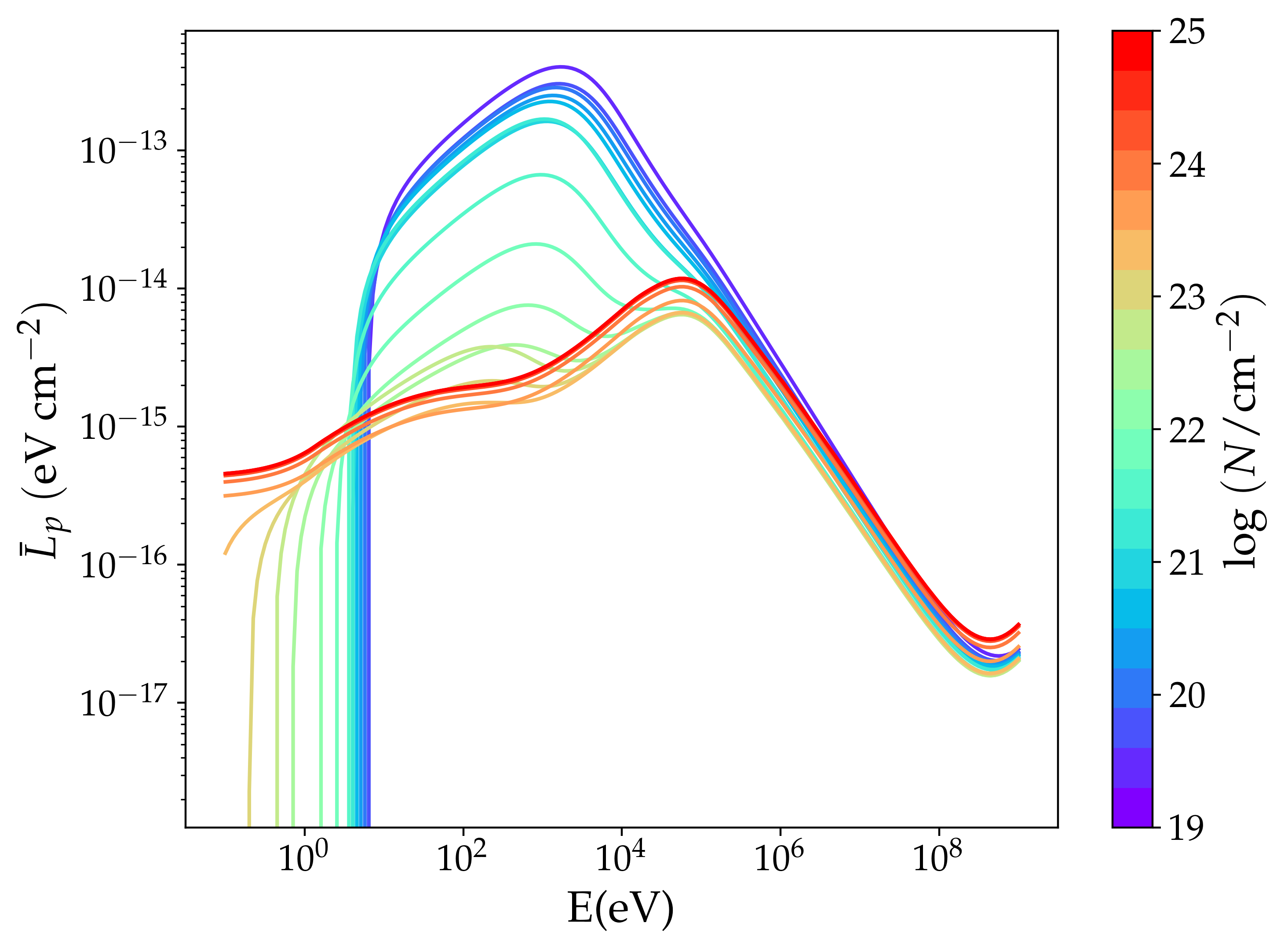}
    \caption{Mean loss functions of protons}
    \label{fig:proton loss function}
\end{subfigure}
\begin{subfigure}{.5\textwidth}
  \centering
    \includegraphics[width=1\linewidth]{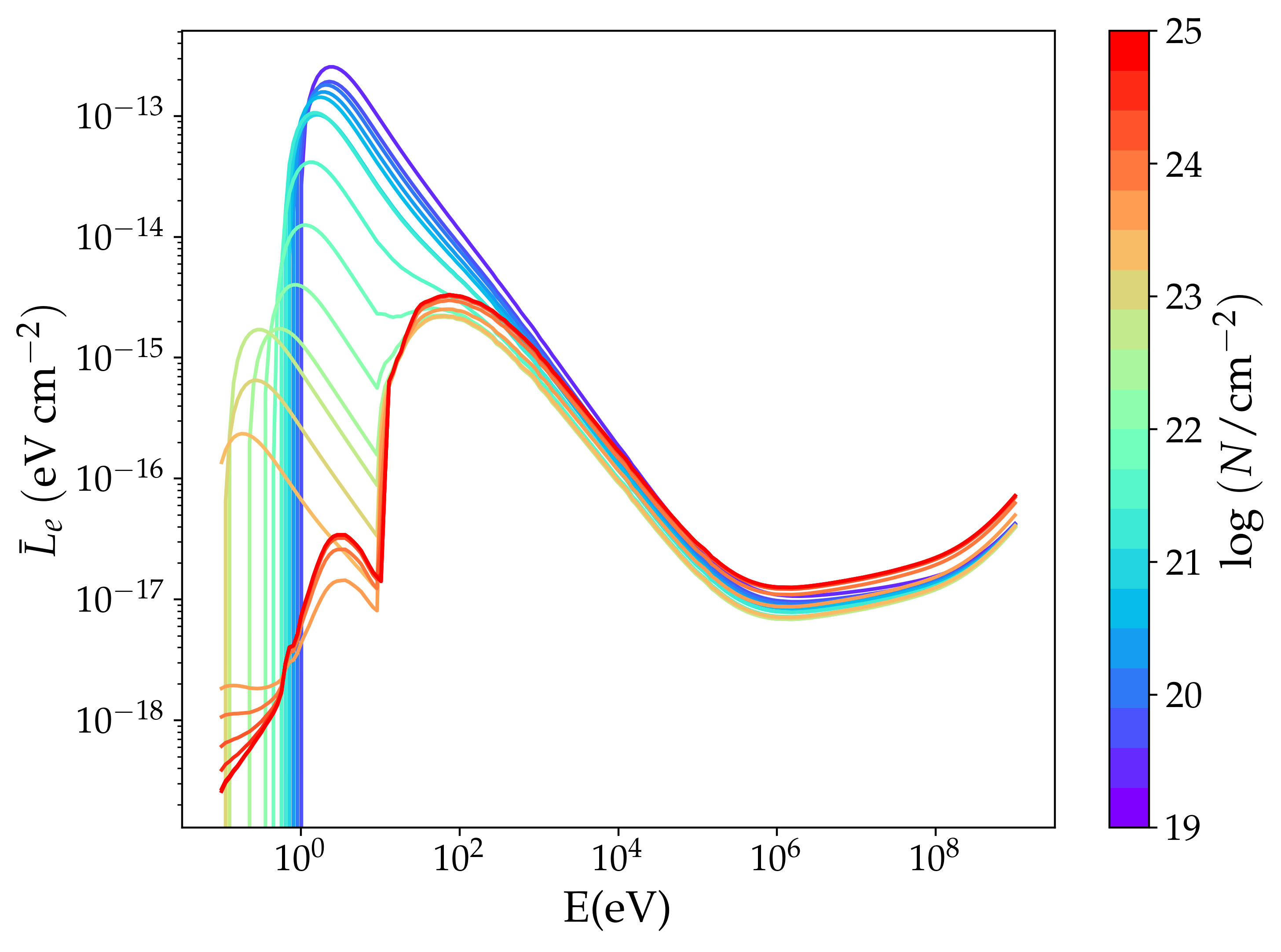}
    \caption{Mean loss functions of electrons}
    \label{fig:electron loss function}
\end{subfigure} 
\caption{
Plots of the mean loss functions vs. particle energy for protons (Fig. a) and electrons (Fig. b). We plot twenty loss functions at column densities N, ranging from $10^{19} \rm cm^{-2}$ (blue curve) to $10^{25} \rm cm^{-2}$ (red curve).}
\label{fig:loss functions}
\end{figure}

Then the calculation for $E(s)$ results from
\begin{equation}
-\int_{E_0}^{E(s)} \frac{\dd E}{  \bar{L}(E',s)} = \int_0^s \sum_i \dv{N_i}{s'} \dd s' = \int_0^s \dv{N}{s'}  \dd s'.
\end{equation}

\subsubsection{Propagated Flux}
The CSDA is based on two assumptions: (i) energy losses are continuous and (ii) the pitch-angle scattering of particles is negligible. \citet{Padovani18} showed that CSDA is valid for particles propagating up to column densities of about $10^{25}$~cm$^{-2}$. At energies higher than $E^\pi$, the interaction between the medium and the energetic protons leads to the production of pions, which rest mass is not negligible compared to the energy of the protons. To study the propagation of particles at high column densities, the formalism of \citet{Padovani18} should be used instead. In this paper, we will not study the propagation to column densities larger than $10^{25}$. Only a future work considering diffusive propagation would enable us to explore the effect of particles deeper in the disc.

The CSDA being valid, the loss functions used in Fig. \ref{fig:loss functions} fully determines the modifications and attenuation of the flux having crossed a column density $N$.
The relation between the injection flux $j(E_0,0)$ and the propagated flux $j(E,N)$ is,
\begin{equation}
    j(E,N)=j(E_0,0) \frac{L(E_0)}{L(E)}.
    \label{eq:spectre CSDA}
\end{equation}
where $N$ is the column density crossed by the particle. Particles with an initial energy $E_0$ reach an energy $E < E_0$ due to energy losses after propagating across a column density $N$:
\begin{equation}
    N=\int_E^{E_0} \frac{\dd E}{\bar{L}(E)}
    \label{eq: N(L,E)}
    \ .
\end{equation}
Eq. \ref{eq: N(L,E)} gives a relation between $E$ and $E_0$, for a fixed column density. As for the loss function, the relation between $E$ and $E_0$ depends on the particle species and the medium in which the particle is propagating. 

We assume that the energetic particles are isotropically emitted during the reconnection event. The particles enter the disc with non-zero pitch angle. The latter is defined as the angle between the particle velocity and the local magnetic field line. The larger the pitch angle, the longer the particle trajectories around the magnetic field lines. 

In order to evaluate the propagated spectra along a field line, we average the contribution of all the particles with different initial pitch angles $\alpha$ over the solid angle,
\begin{equation}
\begin{split}
j(E,N)&=\frac{\int j(E,N,\alpha) d\Omega_\alpha}{\int d\Omega_\alpha} =  \frac{\int_0^{\pi/2} j(E,N,\alpha) \sin(\alpha) d\alpha }{\int_0^{\pi/2} \sin(\alpha) d\alpha }\\
&=\int_0^{\pi/2} j(E,N,\alpha) \sin(\alpha) d\alpha     
\end{split}
\label{eq: solid angle averaged spectra} 
\end{equation}

\begin{figure}
\begin{subfigure}{.5\textwidth}
\centering
    \includegraphics[width=1\linewidth]{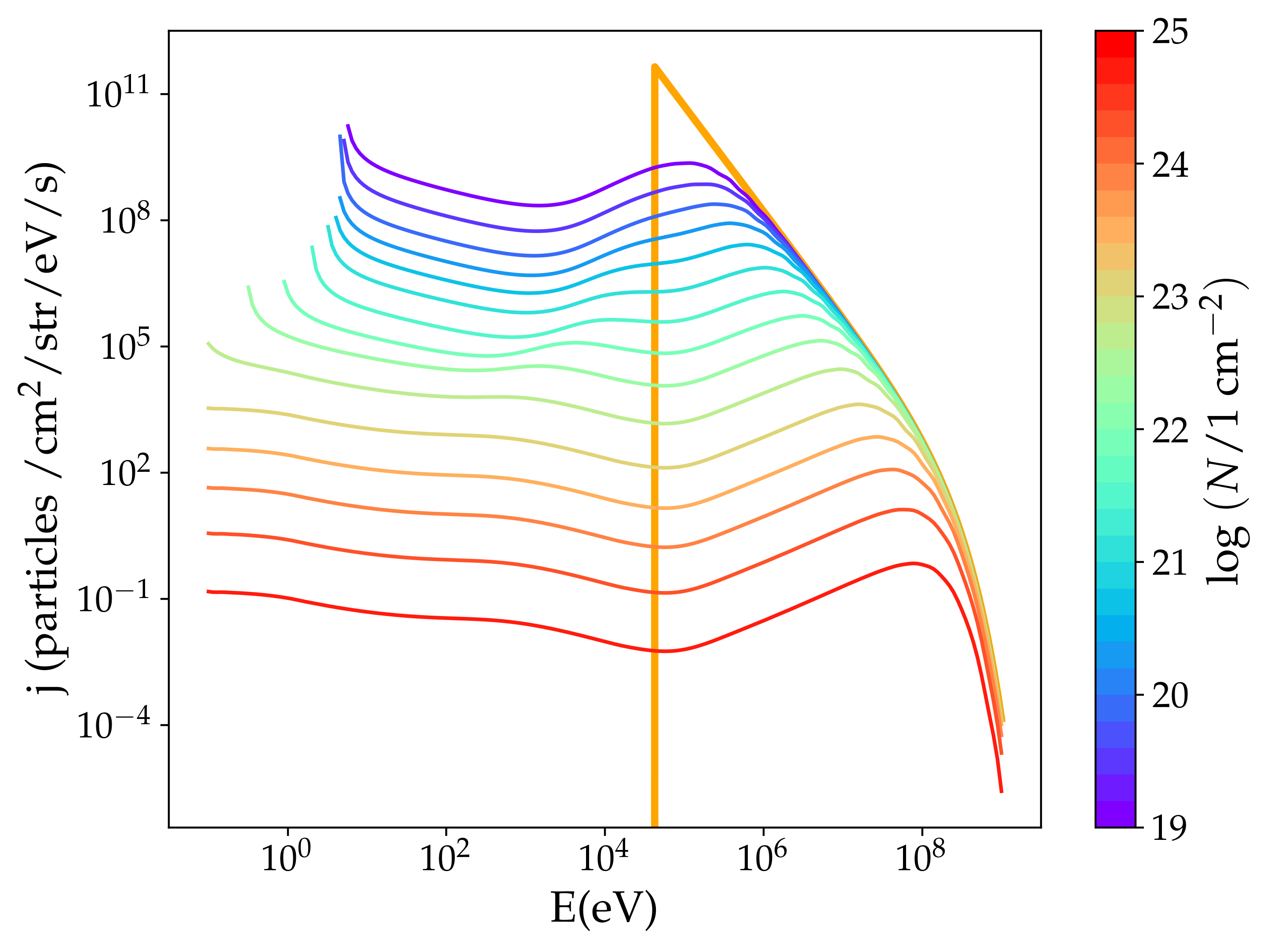}
    \caption{Proton propagated spectra}
    \label{fig:Proton propagated spectrum}  
\end{subfigure}
\begin{subfigure}{.5\textwidth}
\centering
    \includegraphics[width=1\linewidth]{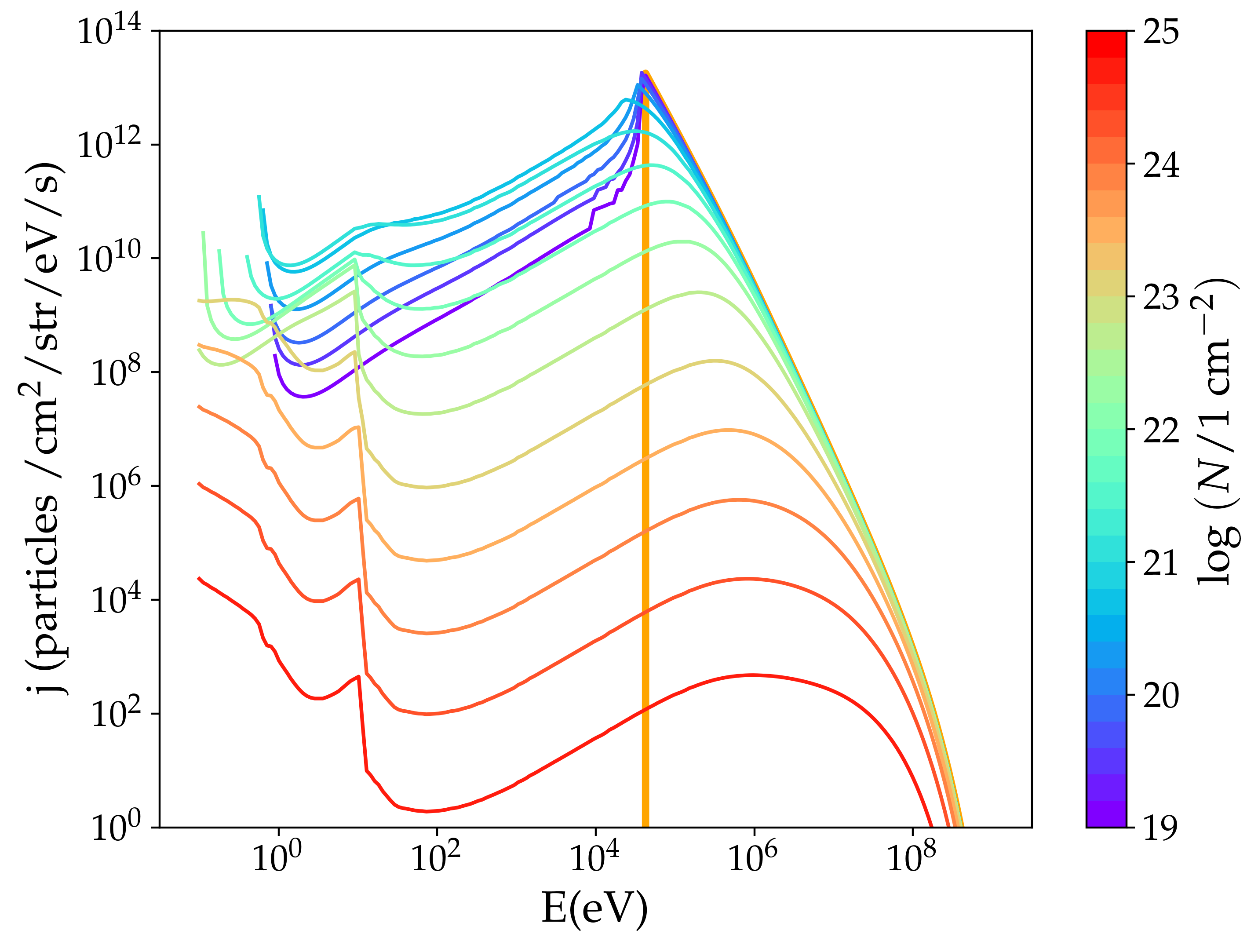}
    \caption{Electron propagated spectra}
    \label{fig:Electron propagated spectrum}  
\end{subfigure} 
\caption{Non-thermal particle spectra at different column densities of the fiducial flare. Fifteen curves are displayed on six decades, equally spaced on the logarithmic scale. The thick orange line plots the injection flux, it is not altered by the medium, yet.}
\label{fig:Propagated spectrum}
\end{figure}
Figure \ref{fig:Propagated spectrum} shows an example of the evolution of the non-thermal particle spectra at different depths in the disc for a flare occurring at a distance $R=0.1~\rm au$ from the star. The orange line in both panels shows the injection flux at the magnetic reconnection site with $\delta=3$.
Comparing Fig. \ref{fig:Proton propagated spectrum} and Fig. \ref{fig:Electron propagated spectrum}, we see that while the proton and electron injection spectra are similar $\left(\rm j_{p,0}(E)= \frac{\beta_p(E)}{\beta_e(E)} j_{e,0}\right) $, they differ in intensity and shape because of the different energy loss processes. Globally, at equal depths, proton fluxes are weaker than the corresponding electron fluxes. This can be explained looking at the energy loss functions in Fig. \ref{fig:loss functions}. At energies lower than 1 GeV (as it is the case of particles injected in our model see Fig. \ref{fig:injection spectra}), the proton energy loss function is always higher than the electron energy loss function. Thus, the column density required to stop an electron is always higher than the column density needed to stop a proton. The difference in shape of the spectra is due to the difference in shape of the electron and proton loss functions. Indeed, for the explored range of column densities at low energies the function $E_0(E)$ is constant, so 
\begin{equation}
j_{e/p}(E)\propto \frac{1}{L_{e/p}(E)}\ .
\end{equation}
At low energies we find the reversed shape of the proton and electron loss functions imprinted over their respective spectra, the spectra are dominated by losses.

\section{Results}\label{S:RES}
\subsection{Ionisation rates}

\begin{table}
\caption{Dominant EP reactions in circumstellar discs}
\begin{tabular}{l|c}
  \hline
  reaction & cross section  \\
  \hline
  $\rm p +H_2 \longrightarrow p + H_2^+ + e$ & $\sigma^{\rm ion.}_{\rm p,H_2}$  \\
  $\rm p +H_2 \longrightarrow H + H_2^+ $ & $\sigma^{\rm e.c.}_{\rm p,H_2}$ \\
  \hline
  $\rm p +H \longrightarrow p + H^+ + e$ & $\sigma^{\rm ion.}_{\rm p,H}$  \\
  $\rm p +H \longrightarrow H + H^+ $ & $\sigma^{\rm e.c.}_{\rm p,H}$ \\
  \hline
  $\rm e+ H_2 \longrightarrow e+H_2^+ +e$ & $\sigma^{\rm ion.}_{\rm e,H_2}$\\
  \hline
  $\rm e+ H \longrightarrow e+H^+ +e$ & $\sigma^{\rm ion.}_{\rm e,H}$\\
  \hline
  
\end{tabular}
\label{table:EP reactions}
\end{table}
The energetic electrons and protons injected into the disc impact the atoms and the molecules in the disc and produce secondary ions and electrons. In Tab. \ref{table:EP reactions} we list the ionisation reactions involving H and $\rm H_2$ targets and the energetic particles (protons and electrons) considered in our model. For reactions involving protons, in addition to ionisation, we take into account the ionisation by electron capture $\sigma^{e.c}$. Electron capture is the dominant ionisation reaction at low energies. Double ionisation, producing two H$^+$ atoms, is a negligible process \citep{Padovani2009}. We have only considered the dominant ion production channel via $\rm H_2^+$. The production rate of ions $i=\rm H,H_2$ per particles is given in \citet{Padovani2009},
\begin{align}
\zeta_i(N)= &  {2 \pi \sum_{k= e,p} \int_{I}^{E_{\rm max}}\left(  j_{ k}(E,N)+2j_{e}^{\rm sec.}(E,N)\right)\sigma^{\rm ion.}_{ k,i}(E)\dd E} \nonumber \\&+ 2 \pi \int_{0}^{E_{\rm max}} {j_{p}(E,N)\sigma^{\rm e.c}_{p,i}(E)\dd E},
\label{eq:ionisation rate}
\end{align}
where $j_k(E)$ is the propagated flux, which corresponds to the number of particles of species $k$ per unit time, area, solid angle and energy, calculated in Eq. \eqref{eq: solid angle averaged spectra}. 

\begin{figure}
\begin{subfigure}{.5\textwidth}
    \centering
\includegraphics[width=1\linewidth]{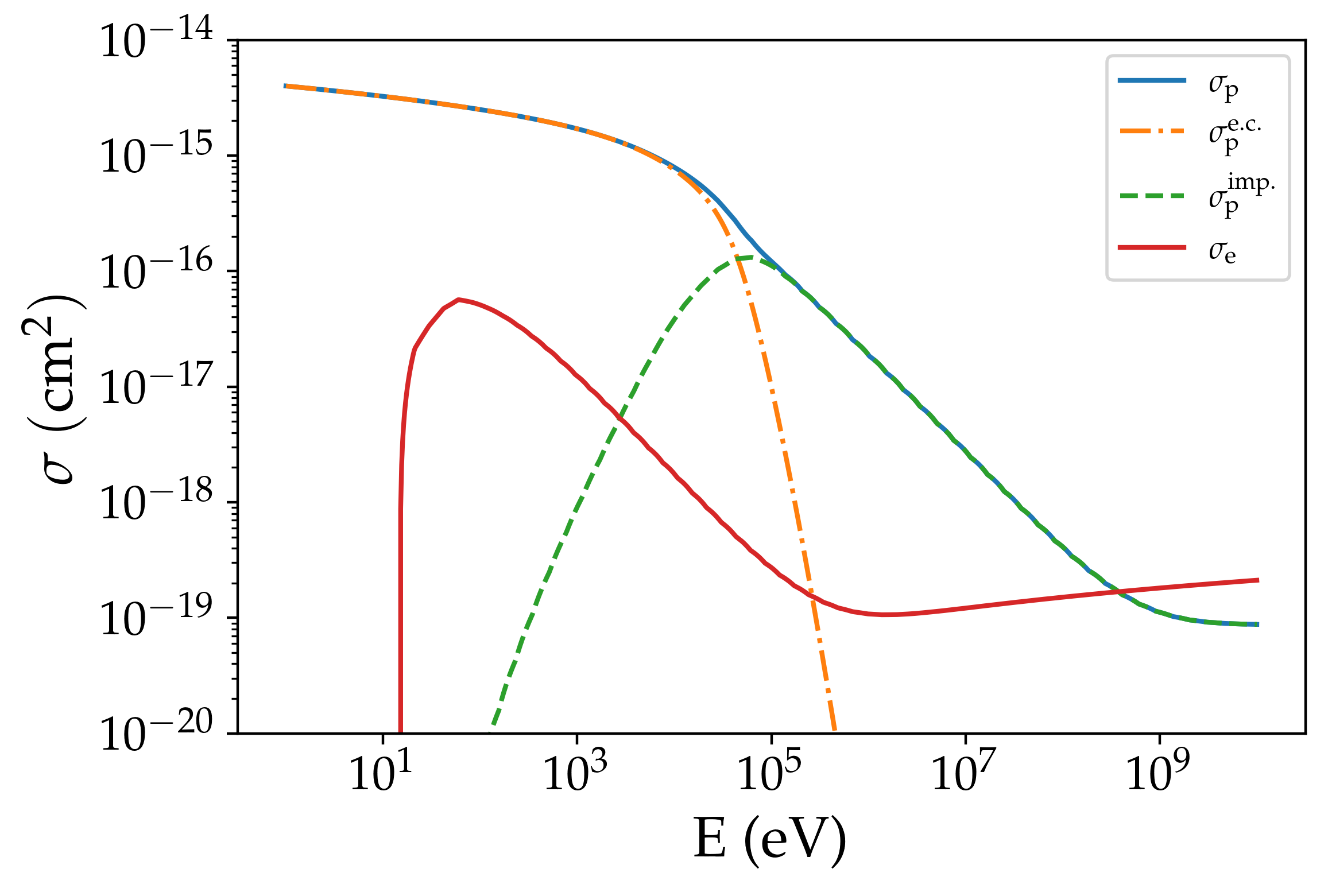} 
    \caption{}
    \label{fig:ionisation cross section H}
\end{subfigure}
\begin{subfigure}{.5\textwidth}
    \centering
\includegraphics[width=1\linewidth]{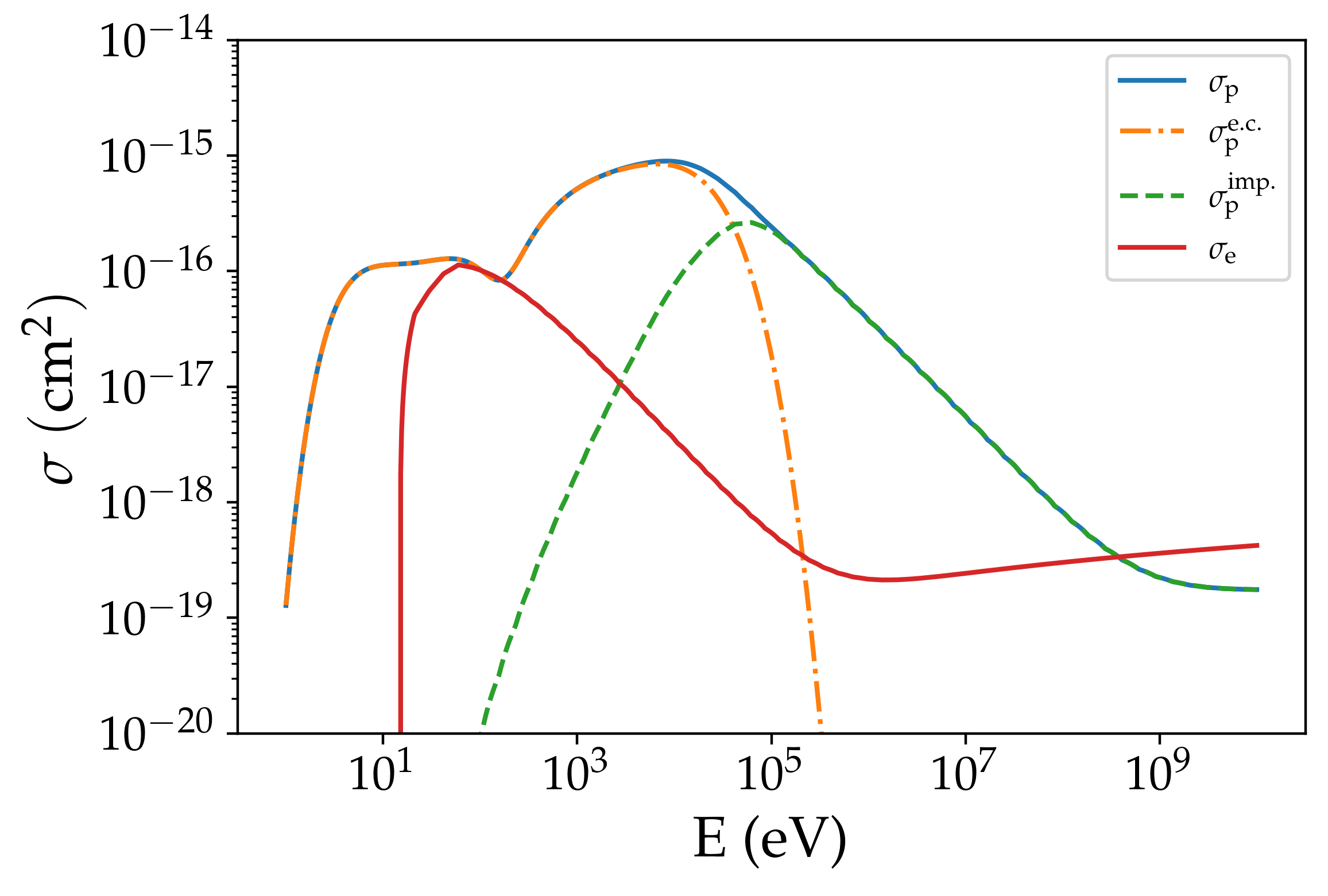}
    \caption{}
    \label{fig:ionisation cross section H2}
\end{subfigure}
\caption{Panel (a) shows ionisation cross sections in atomic hydrogen. Panel (b) shows  ionisation cross sections in molecular hydrogen. The orange dot-dashed lines show the electron capture ionisation cross section by protons ($\sigma_{\rm p}^{\rm e.c}$), the green dashed one show the impact ionisation cross section by proton ($\sigma_{\rm p}^{\rm ion.}$). The blue solid lines show the total ionisation cross section by protons ($\sigma_{\rm p}$). The red solid lines show the total ionisation cross section by electrons ($\sigma_{\rm e}$). In atomic and molecular hydrogen, $\sigma_{\rm e}$ is based on \citet{kim2000extension}. In atomic hydrogen $\sigma^{\rm e.c}_{\rm p,H}$ is based on \citet{janev1993atomic} while in molecular hydrogen $\sigma^{\rm e.c}_{\rm p,H_2}$ is based on Padovani in prep.
In molecular hydrogen $\sigma^{\rm ion.}_{\rm p}$ is based on \citet{krause2015crime}, while  $\sigma^{\rm ion.}_{\rm p,H}$ is assumed to be half of $\sigma^{\rm ion.}_{\rm p,H_2}$.}
\label{fig:ionisation cross section}
\end{figure}
The ionisation cross sections by electrons and protons used in this work are plotted in Fig.\ref{fig:ionisation cross section}. Here, $I$ is the ionisation potential, $ I(\rm{H_2}) = 15.44 ~\rm{eV}$ and $I(\rm{H})=13.60~ {\rm eV}$. In the reactions presented in Tab. \ref{table:EP reactions} electrons are present among the products. These electrons are energetic enough to ionise $\rm H$ and $\rm H_2$ again. The mean free path of the secondary electrons being sufficiently short \citep{Padovani2009}, we can consider that the ionisation by secondary particles is local. With this assumption, we use the expression for the secondary electron flux of \citet{ivlev2015interstellar}, see \citet{ivlev2021rigorous} for a more rigorous calculation of the secondary electron flux.
\begin{equation}
j^{\rm sec.}_{ e}(E) \approx  \frac{ E}{L_{e}(E)} \int_{I+E}^\infty \dd E' j_k(E') \diff{\sigma^{\rm ion.}_k}{E}(E,E').
\end{equation}
Here, $j_k $ is the primary particle flux of $k$, $\diff{\sigma^{\rm ion.}_k}{E}(E,E')$ is the differential ionisation cross section. We use \citet{kim2000extension} for the differential ionisation cross section of electrons and \citet{krause2015crime} for the one of protons. $L_{\rm e}(E)$ is the energy loss function of electrons.

In Eq. \eqref{eq:ionisation rate} the $2\pi$ factor in front of the primary fluxes takes into account that the particles in the primary flux only come from above. Ionisation by secondary electrons, on the other hand, is a local and isotropic process, so the effect of the secondary electron flux is taken into account over $4\pi \rm sr$. 
We then compute a total ionisation rate per hydrogen atom $\zeta$.
\begin{equation}
    \zeta=\zeta_{\rm H}+\frac{1}{2}\zeta_{\rm H_2}.
\end{equation}
\begin{figure}
\centering
\includegraphics[width=1\linewidth]{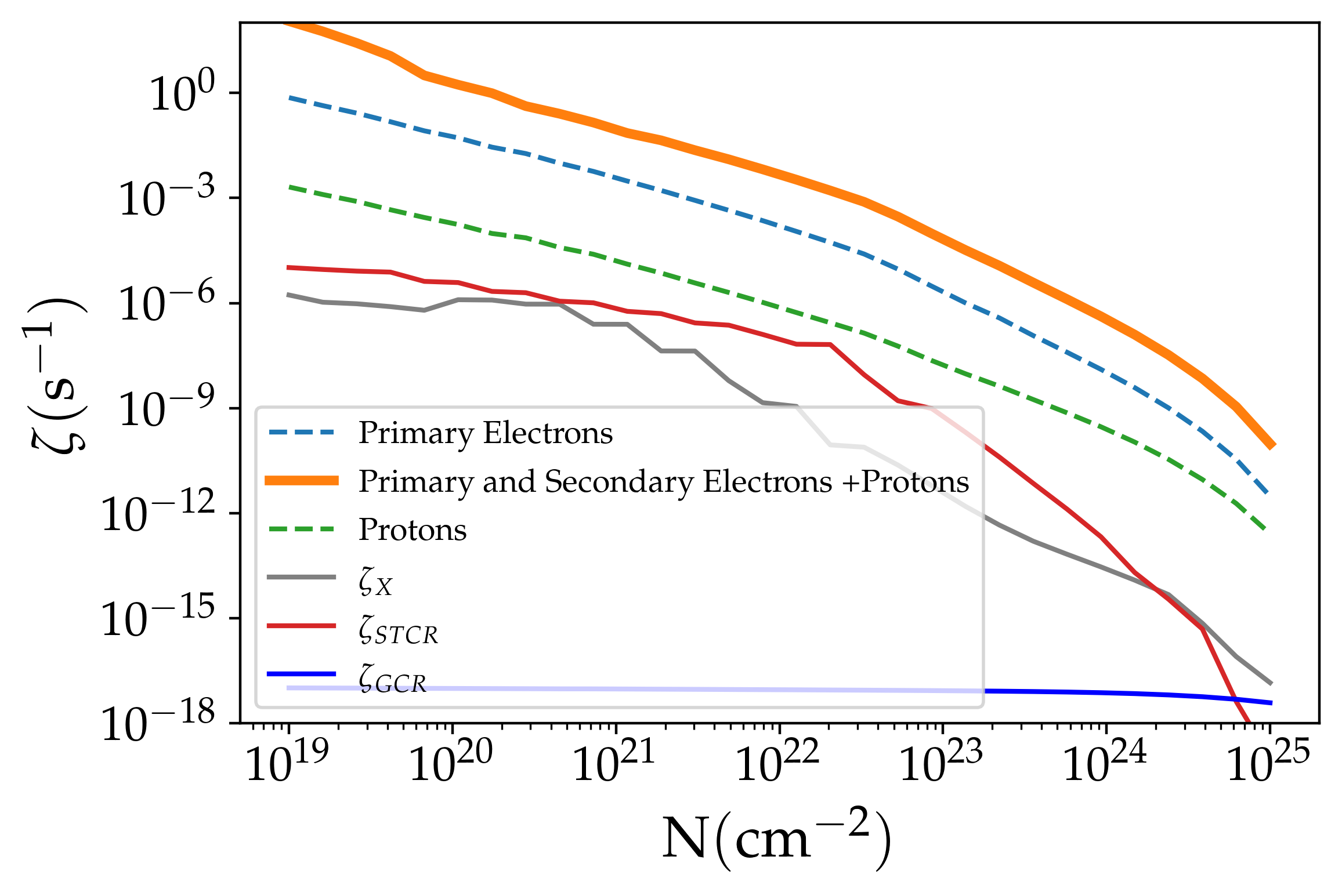} 
\caption{Ionisation rate as a function of the column density accumulated from the flare position. Contributions from primary electrons (dashed blue line), protons (dashed green line), and total including secondary electrons (solid orange line) are plotted separately. These rates are computed for the spectra shown in Fig. \ref{fig:Propagated spectrum}. The solid grey, red and blue lines show the ionisation rate of stellar X-rays, stellar CRs and galactic CRs, respectively, computed by \textsc{ProDiMo}.}
\label{fig:ionisation rates delta=3}
\end{figure}
Fig. \ref{fig:ionisation rates delta=3} shows the ionisation rates produced by the particle spectra of Fig. \ref{fig:Propagated spectrum} as function of the total column density ($N=N_{\rm H}+2N_{\rm H_2}+N_{\rm He}$) and the ionisation produced by stellar X-rays and external GCRs for reference. 
They correspond to the ionisation by a flare occurring at a distance $R=0.1$ au from the star, just above the surface of the disc marked by the yellow line in Fig. \ref{fig:gas distribution}. This flare is injecting non-thermal particles following a power-law flux of index $\delta=3$.
As already mentioned, we do not extend the ionisation rate calculations to column densities higher than $10^{25} \rm cm^{-2}$ since beyond that limit the CSDA is no more valid. We postpone the case of high column densities to a forthcoming work. 

Figure \ref{fig:ionisation rates delta=3} shows that primary electron ionisation is dominant over proton ionisation. This is because in the disc, proton losses are stronger than electron losses (see Fig. \ref{fig:loss functions}), so protons are more attenuated than the electrons. Besides, the injected electron flux is higher than the injected proton flux by a factor $v_e/v_p$ (Eq. \ref{eq:flux proton flux electron}). Most importantly, we see that secondary electrons are the dominant source of ionisation in the disc. Secondary electrons are ejected at energies closer to the maximum of the ionisation cross-section by electrons, at about a hundred electron-volts.

Ionisation rates by stellar X-rays and GCRs as a function of the crossed column density are also plotted in Fig. \ref{fig:ionisation rates delta=3}. It is clear that the ionisation due to particles produced during magnetic reconnection is dominant. However, energetic particles are tight to magnetic field lines and ionise the disc locally. For a better estimation of the ionisation level, the calculations should sample the flare luminosity and location over time. This aspect is further discussed in Sect. \ref{S:DIS}. 

\subsection{A parametric study} 
\label{sec: Parametric study}
In the previous section, we presented the ionisation rates calculated for a reference magnetic field and particle emission model. This fiducial model is based on simple assumptions, such as a vertical magnetic field and a power law injection flux with an index $\delta =3$. The normalisation of the fiducial flux is set by the thermal particle density of a typical flux of temperature $10^6 \rm K$ \citep{getman2008a}. In this section, we conduct a parametric study for these different parameters complemented by a study of the effect of the flare position.

\subsubsection{Effect of spectral index variation}\label{S:IND}
We consider the particle spectra of Fig. \ref{fig:injection spectra} and compute the corresponding ionisation rates in Fig. \ref{fig:ionisation rates delta=[2,8]}. All injection spectra follow a power law with indices $\delta$ ranging from top to bottom from 2 to 8, they are plotted as a function of column density $N$. Harder spectra (smaller $\delta$) produce higher ionisation rates since particles with higher energy act as a reservoir for ionisation. This trend becomes more pronounced at larger column densities.
\begin{figure}
    \centering
\includegraphics[width=1\linewidth]{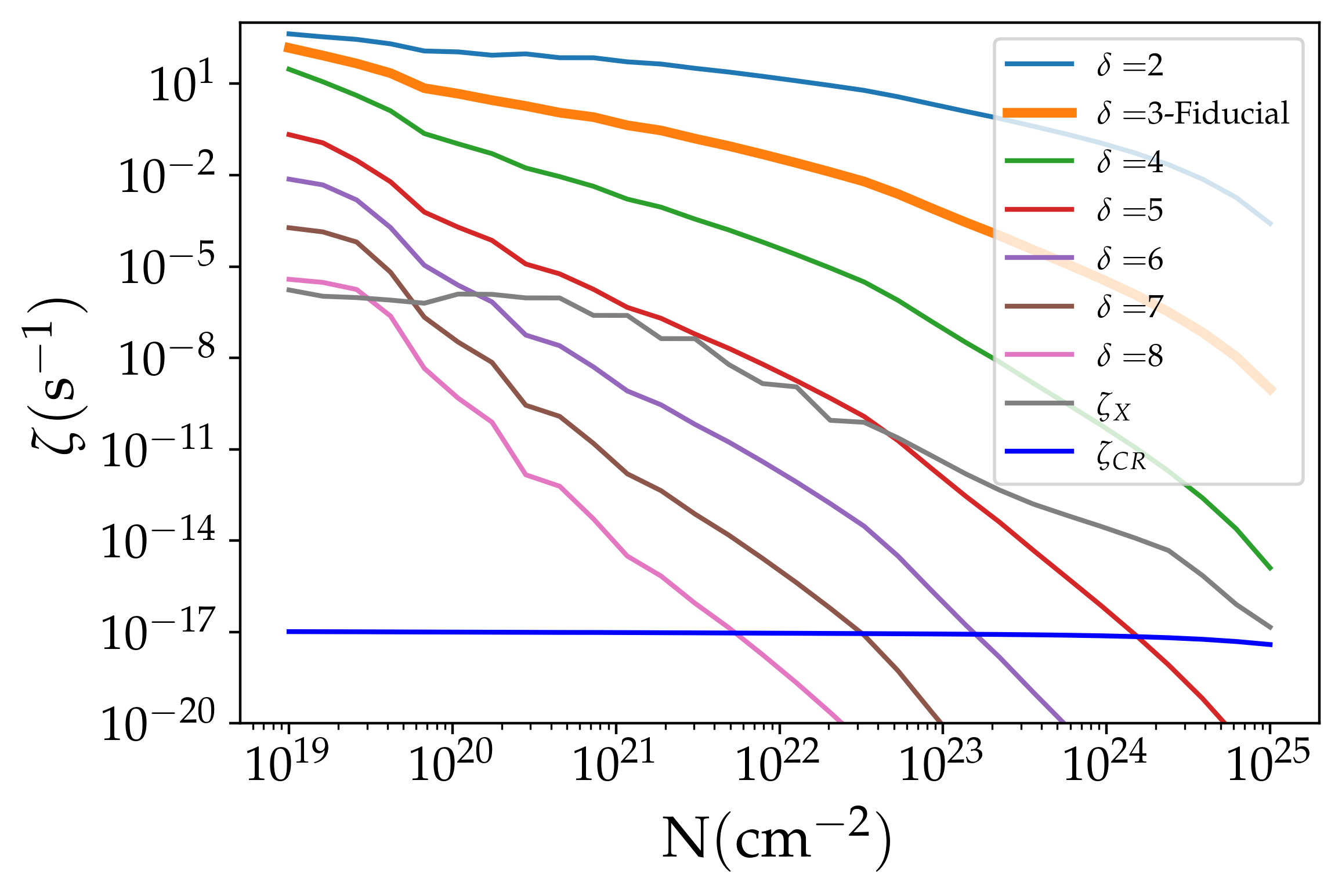} 
    \caption{Ionisation rate as a function of the crossed column density starting from the flare location. This graph shows the ionisation rates produced by energetic particles from a $1 ~ \rm MK$ flare occurring at a distance $R=0.1 ~ \rm au$ from the central star with different indices $\delta$, ranging from top to bottom from 2 to 8. The solid lines are the sum of the proton and electron contributions. The solid grey and blue lines show the ionisation rate of a $1 ~\rm MK$ stellar X-rays flare and galactic CRs computed by \textsc{ProDiMo}.}
    \label{fig:ionisation rates delta=[2,8]}
\end{figure}

\subsubsection{Effect of the magnetic field configuration}\label{S:MFC}
\paragraph{The case of a pure poloidal magnetic field}
The vertical magnetic configuration of the fiducial case (VMF) is a simplistic magnetic configuration. We now test the influence on ionisation rates of more realistic magnetic field models. We examine a hyperbolic magnetic field (HMF) configuration used in standard magnetocentrifugal ejection models and a quartic magnetic field (QMF) configuration constructed to account for a more efficient accretion on the outer layers and the possible formation of a dead zone in the equatorial plane.

\begin{figure}

    \centering
\includegraphics[width=1\linewidth]{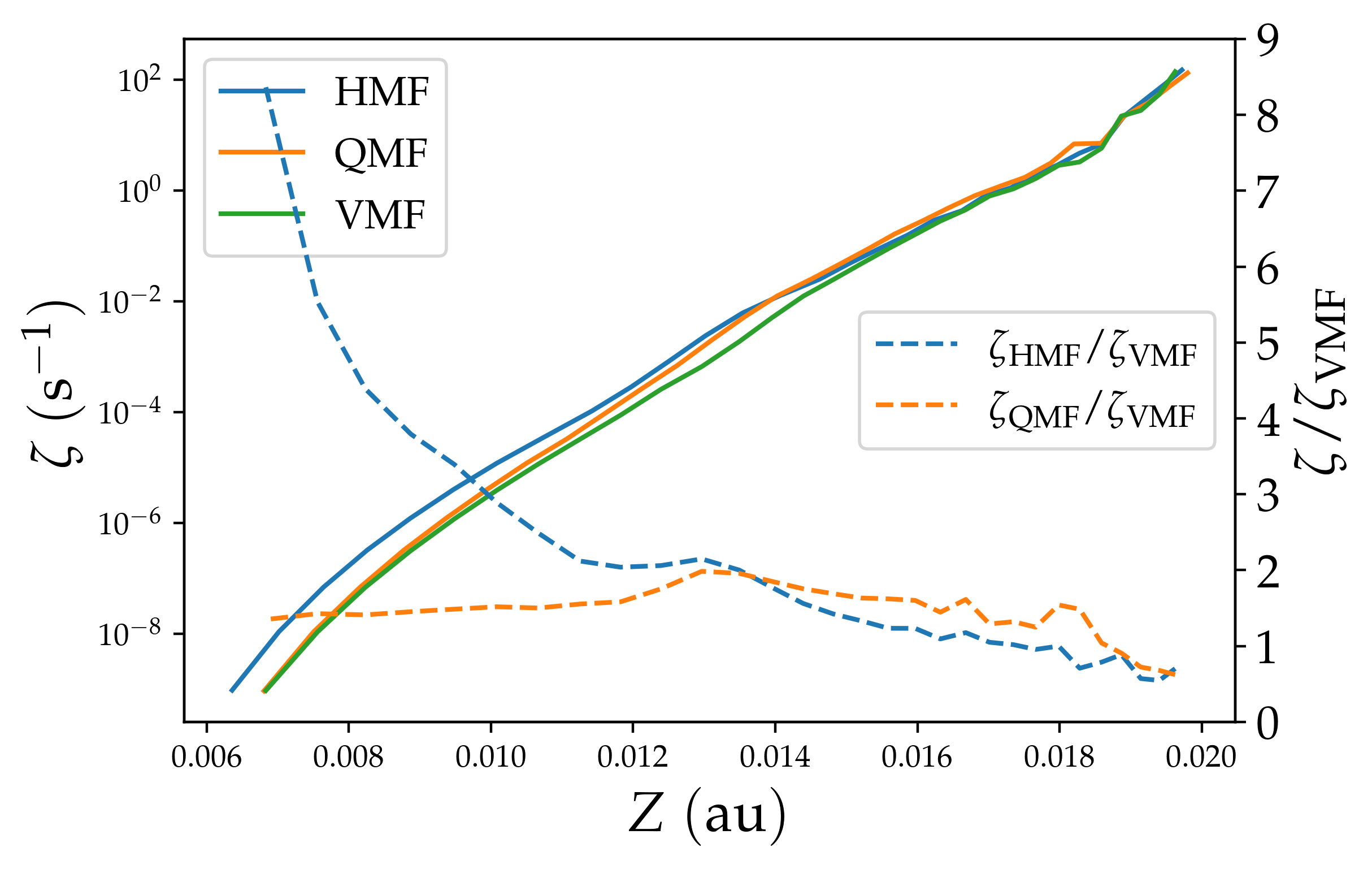} 
\caption{Left vertical axis, plot of the ionisation rate as a function of altitude, above the disc equatorial plane. The EPs are produced by a fiducial flare.
The solid green line represents the propagation along a vertical magnetic field configuration (VMF-fiducial case), the solid blue line along the hyperbolic magnetic field configuration (HMF) and the solid orange line along the quartic magnetic field configuration (QMF). 
Right vertical axis, plot of the ionisation rate of the HMF and QMF configurations relative to the VMF reference configuration, respectively the dashed blue and orange lines.
The altitude range corresponds to the column density range considered, from $10^{19} ~\rm cm^{-2} (Z \approx 0.020 ~ \rm au )$ to $10^{25} ~\rm cm^{-2} (Z \approx 0.007 ~ \rm au)$. }
\label{fig: Magnetic configurations}
\end{figure}

Even though these configurations transport the particles to different regions of the disc, the relative abundances of the chemical species as a function of the column density are very similar. However, when ionisation rates are plotted as a function of the altitude above the equatorial plane of the disc, there is a significant difference between the three configurations (see Fig. \ref{fig: Magnetic configurations}). This difference comes from the fact that, at the same altitude, the column density explored by the particles following QMF and HMF are lower than that explored by VMF, so the QMF and HMF flux are less attenuated and therefore produce higher ionisation rates. This difference is especially significant for the HMF configuration at lower altitudes. The relative ionisation rate ratio shows that the ionisation rate of the HMF configuration is up to almost an order of magnitude higher than the ionisation rate of the VMF configuration (see dashed lines Fig. \ref{fig: Magnetic configurations}.

\paragraph{The role of the toroidal magnetic field}
In our study, we investigate the influence of the presence of a toroidal component of the magnetic field on the ionisation in the disc. This component has two effects. First, it forces the non-thermal particles produced by the flare to explore different regions of the disc, thus different column densities. The larger the toroidal component, the larger the column density to cross to reach a given column density in the inner disc. Then, the larger is the crossed column density, the more attenuated is the propagated particle flux. Consequently, by increasing the toroidal component, the ionisation rate drops as a function of altitude. This is a geometrical effect.
Second, the toroidal component plays an essential role in the process of particle acceleration by magnetic reconnection. Many studies have shown that a guide magnetic field, perpendicular to the plane where the magnetic reconnection is occurring, reduces the efficiency of particle acceleration. \citet{che2020electron} showed that the spectral index is determined by the ratio, $b_g$, of the guiding magnetic field, $B_g$, to the asymptotic magnetic field, $B_0$. Simulations by \citet{che2020electron} and \citet{arnold2021electron} give similar expressions for the power law index,
\begin{equation}
    \delta \approx 2.5+4 b_g^2 \ .
    \label{eq:index-guide field}
\end{equation}
Magnetic reconnection is an intrinsic 3D turbulent process and the notion of guiding field may be a bit artificial, but we consider that reconnecting fields in our {\it asymmetric} configuration mostly occur between the dipolar stellar magnetic field and the poloidal disc magnetic field. This configuration implies that the toroidal magnetic field can be approximated as a guiding field and hence we adopt $b_g$ the ratio of toroidal to poloidal magnetic components as a proxy of this guiding field. Using Eq. (\ref{eq:index-guide field}) we can simply evaluate the impact of the toroidal magnetic field over the disc ionisation process. However, we have to acknowledge the differences between a realistic environment with respect to numerical experiments from which the above relation between $\delta$ and $b_g$ is derived. This probably leads to an overestimation of the effective value for $\delta$. 

\begin{figure}
    \begin{subfigure}{.5\textwidth}

    \centering
    \includegraphics[width=1\linewidth]{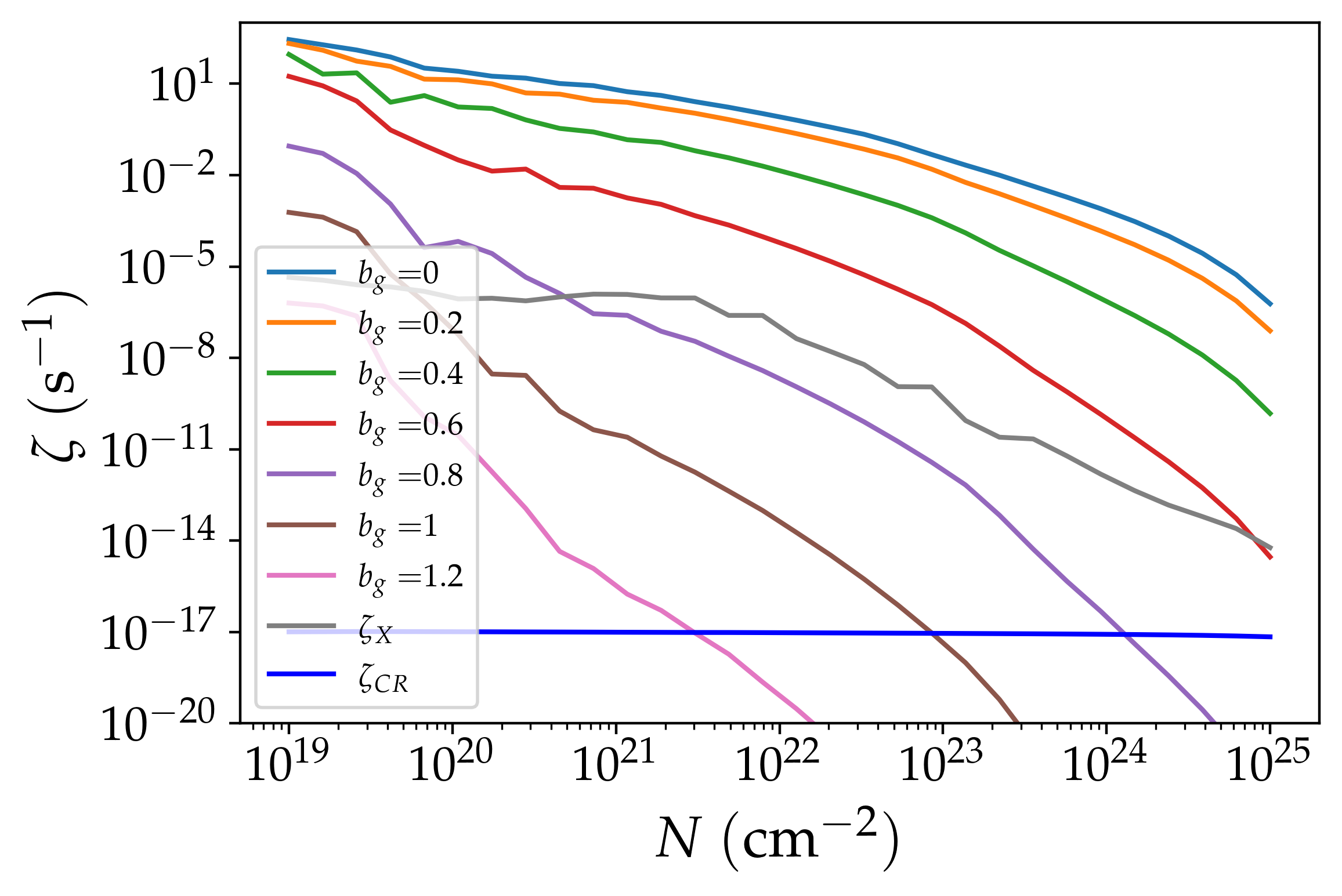}
    \caption{Ionisation rates as a function of the crossed column density from the flare location.}
    \end{subfigure}
    \begin{subfigure}{.5\textwidth}
     \centering
    \includegraphics[width=1\linewidth]{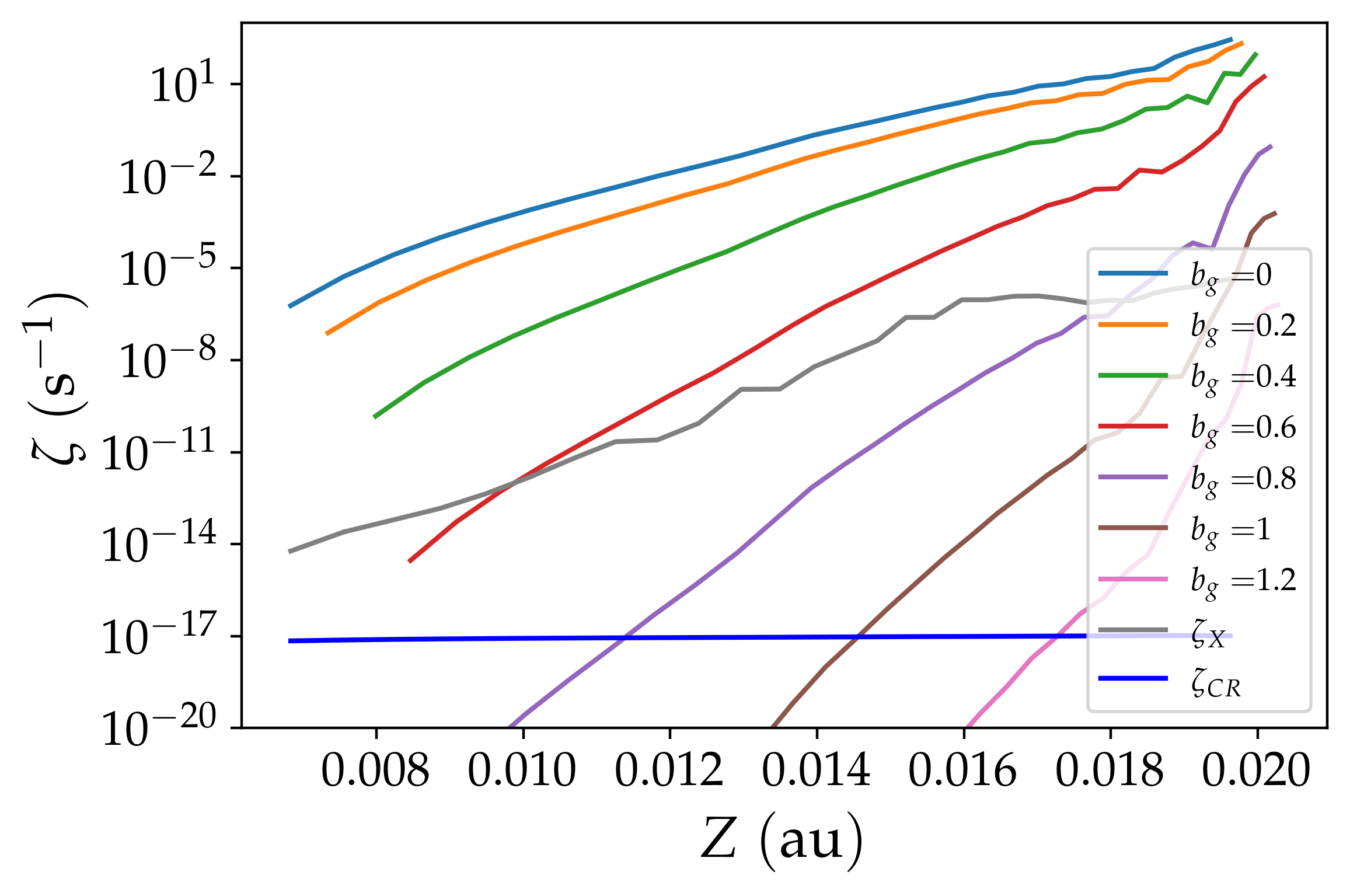}
    \caption{Ionisation rates as a function of altitude above the disc mid-plane.}
    \end{subfigure}
    \caption{Ionisation rates produced by energetic particles propagating along magnetic fields for different toroidal to poloidal magnetic field ratio. For each ratio, the index of the injection flux of the EP is given by Eq. \eqref{eq:index-guide field}. The solid grey and blue lines show the ionisation rate of stellar X-rays and galactic CRs computed by \textsc{ProDiMo}.}
    \label{fig:ionisation toroidal}
\end{figure}

In Fig. \ref{fig:ionisation toroidal}, we plot the ionisation rates for a set of toroidal to poloidal magnetic field ratio $b_g$ ranging from 0 to 1.2. Panel (a), which shows the ionisation as a function of the column density, shows the effect of the toroidal component only through its influence on the spectral index of the injection flux (Eq. \ref{eq:index-guide field}). To highlight the impact of the toroidal field on the ionisation rate due to both the geometry and the spectral index, we plot the ionisation rate for a set of values of $b_g$ as a function of the altitude in panel b of Fig. \ref{fig:ionisation toroidal}. 

\subsubsection{Effect of the flare temperature}
The flare temperature has a substantial effect on the ionisation rate in discs. We rely on the observations of \citet{getman2008b} to obtain a range of temperatures, from 1 MK to a few hundreds of MK. In the following we consider flares at three different temperatures, 1, 10, and 100 MK. The temperature of the flare has globally two consequences. The first one is due to X-ray emission, as discussed in Sec. \ref{section:X-ray flare model}. Hot flares have higher X-ray luminosities, and X-ray emission changes the structure of the disc first, as energetic particles are lagging the photons. In our calculations of the ionisation rates by energetic particles, we have used a \textsc{ProDiMo} disc model that takes into account the modification of temperature and the chemical structure of the disc by the X-ray emission corresponding to the temperature of the studied flare. 

\begin{figure}
\begin{subfigure}{.5\textwidth}
    \centering
\includegraphics[width=1\linewidth]{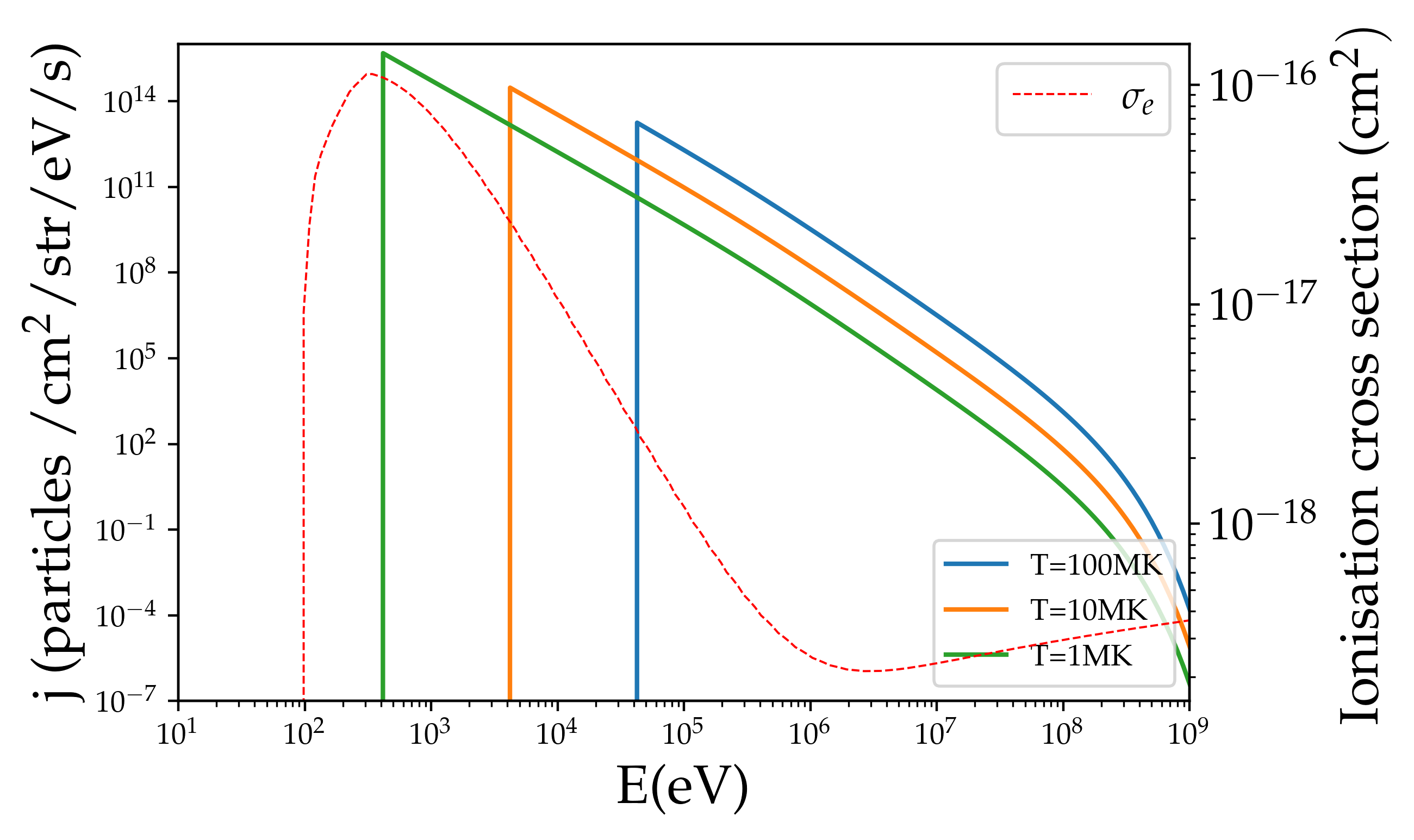} 
    \caption{Electrons}
    \label{fig:Injection temperature cross section Electron}
\end{subfigure}
\begin{subfigure}{.5\textwidth}
    \centering
\includegraphics[width=1\linewidth]{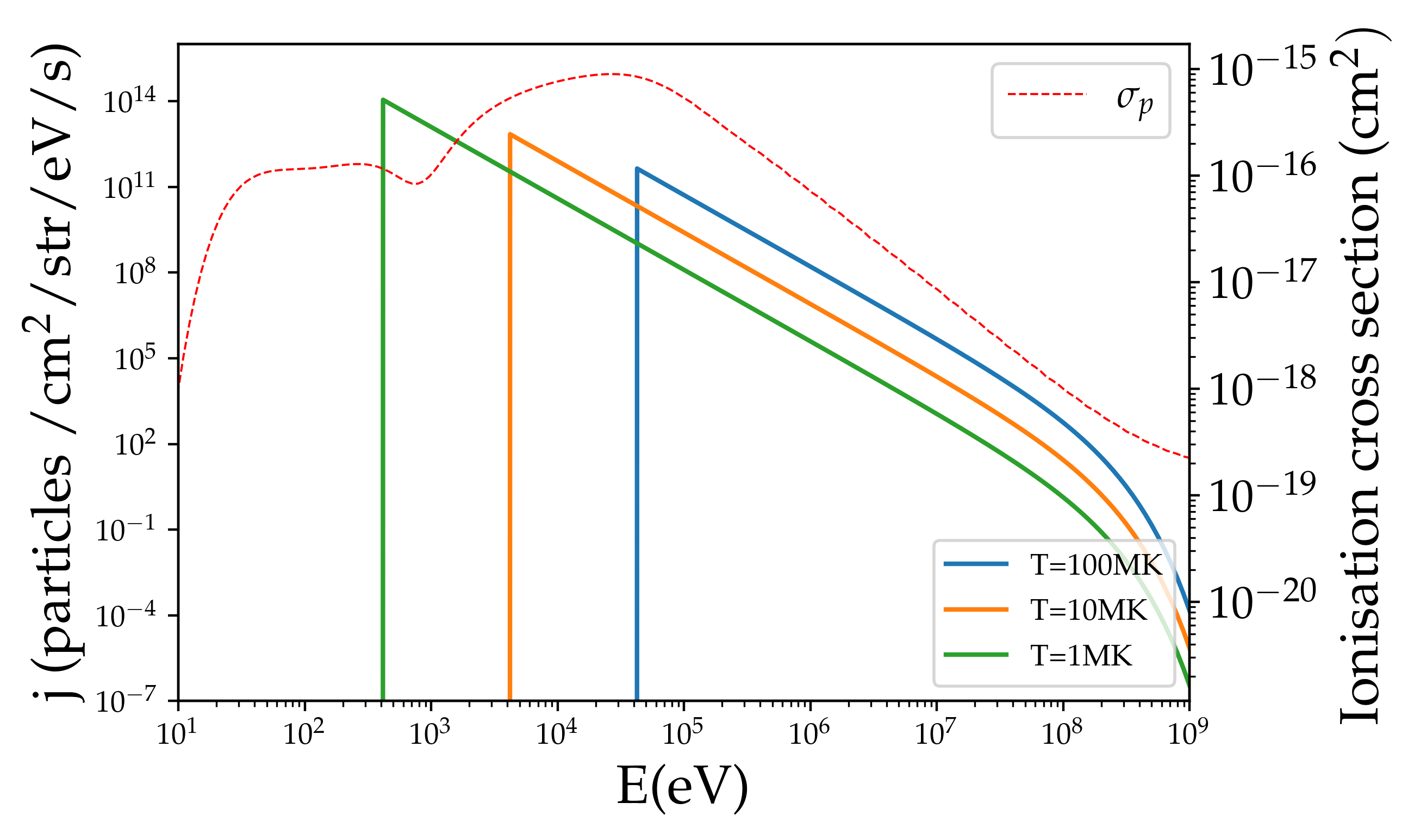}
    \caption{Protons}
    \label{fig:Injection temperature cross section Proton}
\end{subfigure}
\caption{Non-thermal injection spectra for electrons and protons (panel (a) and (b), respectively) for different flare temperatures, 100, 10, and 1 MK (solid blue, orange, and green lines, respectively). We have superimposed the ionisation cross sections of electrons and protons (resp. $\sigma_e$ and $\sigma_p$) on atomic hydrogen in dashed lines.}
\label{fig:Injection temperature cross section}
\end{figure}

The second impact of the flare temperature is linked to the flux of injected non-thermal particles. The temperature changes the normalisation of the injection flux but also their injection energy. The normalisation of the flux to the injection energy decreases at higher temperatures as it can be seen in Eq. \eqref{eq:nNth}, but the injection energy is higher ($E_{\rm c}=3E_{\rm th}$). It follows that the whole injection flux of a high-temperature flare is shifted towards higher energies, see Fig. \ref{fig:Injection temperature cross section}. 

\begin{figure}
    \centering
    \includegraphics[width=1\linewidth]{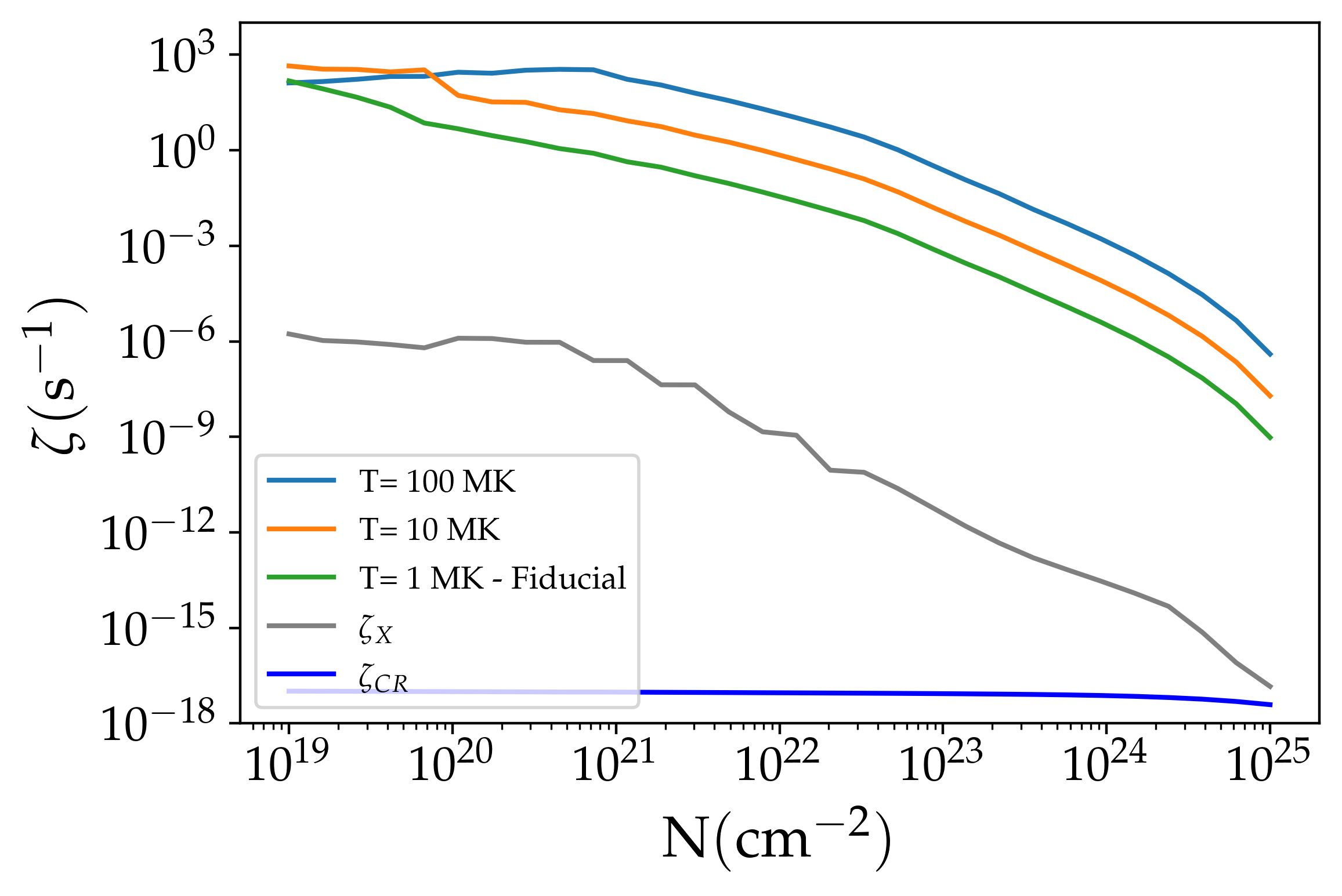}
    \caption{Ionisation rate as a function of the column density accumulated from the flare for three different flare temperatures. The solid grey and deep blue lines show for comparison the ionisation rates of stellar X-rays and GCRs computed by \textsc{ProDiMo}.}
    \label{fig:ionisation rate temperature}
\end{figure}

A low normalisation produces lower ionisation rates whereas a flux shifted towards high energies produces higher ionisation rates. These two parameters have opposite effects. However, it can be seen from Fig. \ref{fig:ionisation rate temperature} that the hotter the flare, the higher the ionisation rate for column density higher than $10^{20} ~\rm cm^{-2}$. At lower column densities for hot flares, the population of particles with energies of the order of $100 ~\rm eV$ (close to the ionisation cross-section by electrons maximum) is not filled yet (see Fig. \ref{fig:Electron propagated spectrum}). Thus at low column densities, hot flares do not ionise as much as cold flares. We see a change in the relative ionisation at $10^{20} ~\rm cm^{-2}$, the flare at $10~\rm MK$ starts to ionise more than the one at $100~\rm MK$ and at $10^{19} ~\rm cm^{-2}$, the $1~\rm MK$ flare ionises more than the $10~\rm MK$ flare. From this, we deduce once again that at high column density, the ionisation rates are driven by high energy particles.

It is important to note that even the coolest flares ($1~ \rm MK$) produce, at least locally, much higher ionisation rates than those produced by X-rays. In a future work, we will take into account temporal effects by considering flares sampled by temperature. Colder flares are expected to be more frequent, so the results shown in Fig.~ \ref{fig:ionisation rate temperature} suggest that these cold flares could have a strong influence on the time-averaged ionisation in the disc.

\subsubsection{Effect of the flare position}
\begin{figure}
\includegraphics[width=1\linewidth]{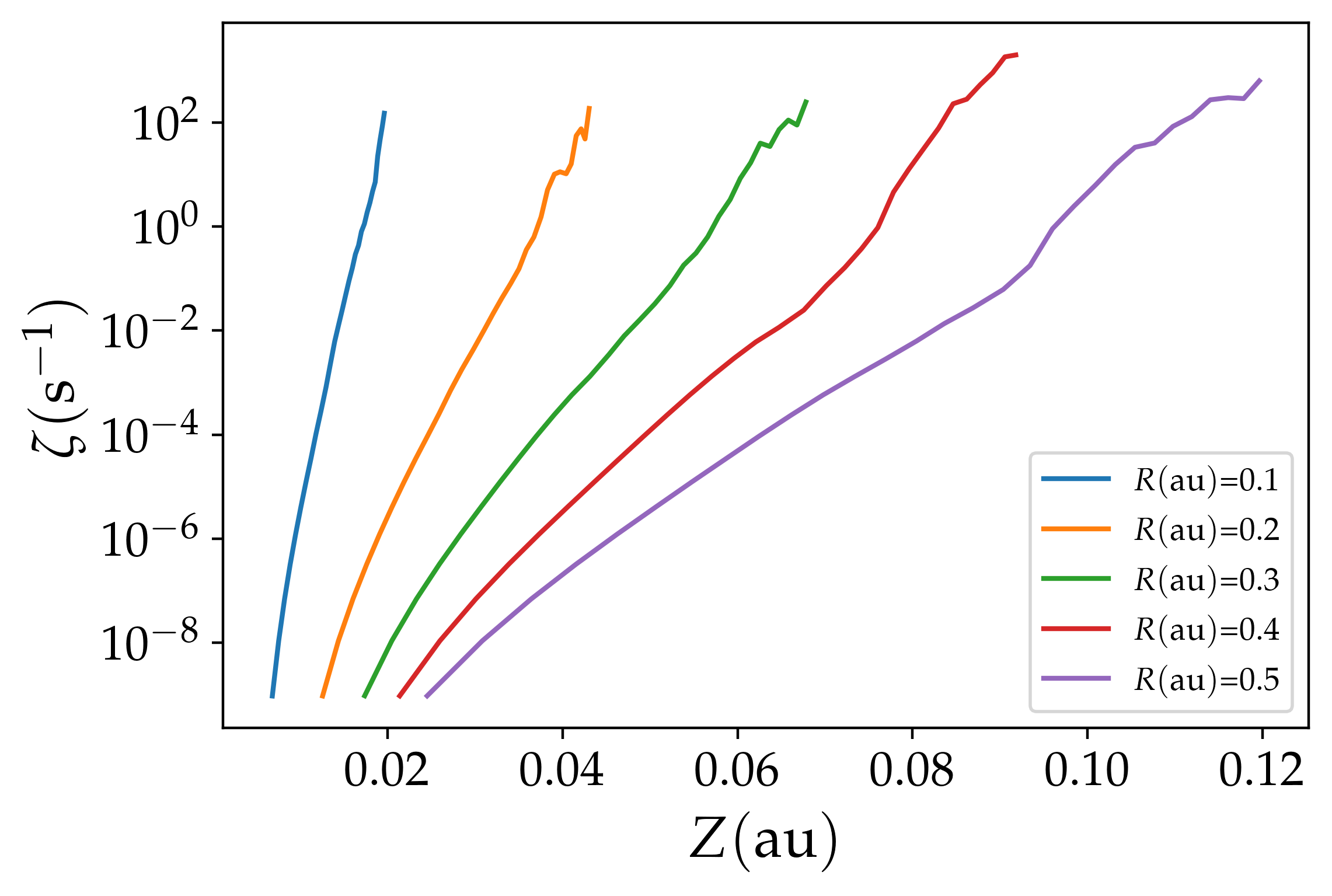}
\caption{ Ionisation rate as a function of the altitude above the disc equatorial plane. The EPs are produced by fiducial flares occurring at 0.1, 0.2, 0.3, 0.4, 0.5 au from the central star. The altitude range corresponds to column densities varying from $10^{19} ~\rm cm^{-2}$ to $10^{25} ~\rm cm^{-2}$. }
\label{fig:ionisation rate radius}
\end{figure}

So far we have considered flares occurring at a distance $R=0.1 ~\rm au$ from the central star. This distance corresponds to the innermost part of the disc that we have access with \textsc{ProDiMo}. The region of the disc closest to the star's magnetosphere is where we expect flares produced by the interaction of the magnetic fields of the star and the disc to be most frequent. Nevertheless, it can be assumed that flares also occur at greater distances from the star. Therefore, we consider the effect on the ionisation of the disc by flares occurring at larger distances from the star. Fig. \ref{fig:ionisation rate radius} shows the ionisation rates produced by flares occurring from $R=0.1 ~ \rm au$ to $0.5 ~ \rm au$. 
We find that the larger the radius, the smaller the ionisation rate gradient. This is directly related to the decrease of the vertical density gradient with increasing radius.

\section{Discussion}\label{S:DIS}

In Sect.\ref{S:RES} we have shown that the injection of non-thermal particles by strong magnetic reconnection events in the close environment of a T Tauri star leads to enhanced ionisation rates in the inner disc region. In the following subsections we mitigate our results by discussing their limitations.

\subsection{Disc model}
In order to calculate the ionisation rate in the inner disc, we take benefit of the very sophisticated circumstellar disc model, \textsc{ProDiMo}. Although \textsc{ProDiMo} is capable of compute the abundances of many chemical elements, we decided to consider only a few chemical species (e, H, H$_2$, He). This approximation is quite acceptable for our calculations since these four species largely dominate in abundance and hence control the energy losses of the particles produced by the flares. As already mentioned, \textsc{ProDiMo} does not include magnetic fields. A solution of magnetic field lines calculated by a 3D MHD code would allow an accurate description of the trajectory of the particles in the disc. But such a code, taking into account the chemical and magnetic dynamics of a circumstellar disc does not exist right now to our knowledge. As such a code is not currently available, we superimposed different magnetic field configurations with typical morphology expected in circumstellar discs on the \textsc{ProDiMo} disc model. Apart from the vertical magnetic field configuration which is not realistic even if it provides us a reference case, we propose two other configurations, namely the QMF and the HMF, which are representative of the magnetic field configuration expected in discs \citep{orlando2011mass, king2007accretion}. For now, we do not aim at using more realistic magnetic configurations, i.e. specifically respecting $\vec{\nabla}\cdot\vec{B}=0$ but rather to evaluate the respective role of poloidal and toroidal components into the propagation of energetic particles. We find that with respect to the VMF case poloidal configurations produce an increase of the ionisation rate by a factor up to 10 because of less column densities traversed at a given altitude whereas the presence of a toroidal magnetic field component produces a decrease of the ionisation rate of a factor up to 600.

\subsection{Radiation model}\label{ss:Radiation model}
\textsc{ProDiMo} is able to calculate a disc structure taking into account the effect of an X-ray field. We have computed the effect of X-rays emitted by a flare to change the structure of the disc before the propagation of the energetic particles. To have the most accurate description of such an effect, X-rays should be located at the flare position. In its current state, \textsc{ProDiMo} only handles a flare located at the central star. Setting X-rays at the flare location would produce a locally enhanced ionisation of the medium, which should modify the chemical distribution and the column density in each elements. We expect the structure of the inner disc to be only lightly sensitive to the X-ray field from the star. Indeed, the region of the disc in which the chemical distribution is strongly modified by the X-ray flare extends over a small column densities ($N <10^{19} ~ \rm cm^{-2}$ ) in the disc. Energetic particles are not sensitive to changes in the chemical distribution in such a small column density range. Therefore, the origin of the X-rays is not expected to have a significant influence on the ionisation rates due to particles emitted by the flare. However if we consider X-rays emitted at the flare location, at a distance $R_{\rm flare}=0.1~\rm au$ from the star, there is less geometrical dilution of the flux. An approximate estimation is an increase of the X-ray flux of a factor $\left(\frac{R_{\rm flare}}{R_*}\right)^2\approx 10^2$. Hence, placing the origin of the X-ray emission at the flare location  produces an increase of the ionisation rate due to the X-rays by the same factor $\approx100$. It appears that even by increasing the X -rays ionisation rate by two orders of magnitude, ionisation by non-thermal particles stay completely dominant, see Fig \ref{fig:ionisation rates delta=3}. This increases may however have to seen as an upper limit. In effect, above density columns about $10^{21}~\rm{cm^{-2}}$ X-rays ionisation rate drop (see Fig  \ref{fig:ionisation rate temperature}). We do not then expect the location of the X-ray to have strong impact over the ionisation at large column density in the disc.

\subsection{Particle injection model}
The purpose of this work is to calculate ionisation rates produced by particles emitted from a magnetic reconnection site at the stellar magnetosphere-accretion disc interface. 
The injection flux depends on several parameters, the particle distribution normalisation, the injection energy, the spectral index and the cut-off energy. In what follows we always consider the VMF configuration.

A variation by a factor 10 in the cut-off energy with all other parameters fixed does not change the ionisation rate by more than a factor of 5 for column densities lower than $10^{25}~\rm cm^{-2}$. A variation of the injection energy by a factor 2 does not produce a variation by more than a factor 8. We have already discussed the effect of the spectral index in Sect. \ref{S:IND}. A change of the index from 3 to 4 (to 2) produces a decrease (an increase, respectively) of the ionisation rate by three orders of magnitude at $N=10^{25}~\rm{cm^2}$. Then, softer indices (larger $\delta$), accounting for the effects of a variation of the injection energy, produce even lower ionisation rates. Finally, a variation of the size of the reconnection length scale $L$ by a factor 10, that is essentially a variation of the normalisation of the injection flux by a factor $L^3$, induces a change by a factor 1000 on the ionisation rates. We hence have quite large uncertainties over the ionisation rates due to the particle distribution normalisation and spectral index. As we shall discuss below, a more realistic approach consists in considering a time-dependent effect of a sample of flares. Given the parameter space that a single flare can explore, it is expected that the ionisation rates produced by flares arising from a distribution can vary by several orders of magnitude. We postpone to a following paper a quantitative analysis of this phenomenon. 

\subsection{Propagation model}
The results presented in this paper predict a very strong contribution of the particles produced by flares on the ionisation of the inner part of T Tauri discs. In the studied column density range, even for flares with the softest particle injection spectra ($\delta =8$), ionisation rates are orders of magnitude higher than those produced by X-rays from flares taking place on the surface of the central star and and by the GCR flux \citep{Rab17}, see Fig. \ref{fig:ionisation rates delta=3}. The contribution of these EPs is expected to be dominant even beyond the maximum column density explored here ($N>10^{25} ~ \rm cm^{-2}$). To validate this hypothesis, it seems crucial to improve our particle propagation model in order to be able to take into account several effects. In particular, at column densities $> 10^{25}~\rm{cm^2}$, the CSDA is not valid anymore and particles start a diffusion process \citep{Padovani18}. This aspect will be treated in a forthcoming work. 

\subsection{On the origin of the X-ray flares}
Since the first X-ray detections of flares in young stars, theoretical models have been developed involving interactions between the magnetic field of the disc and the star. This coupling is especially motivated by its role in removing angular momentum and slowing down the rotation of protostars \citep{bouvier1997angular}. The torque that slows the rotation down is most easily explained by stellar magnetic field lines anchored to the disc. And since the disc is rotating differentially, shearing and reconnection of star-disc field lines are expected. Such flare models from star-disc interaction were calculated by \citet{orlando2011mass} and then by \citet{colombo2019new} in 3D MHD. Our work is also based on these star-disc interaction models, which have many astrophysical motivations. However, recent data from \citet{getman2021b} do not find strong evidences for such interaction processes. The X-ray properties of the flares observed in protostars with and without discs are statistically indistinguishable. Their analysis therefore provide no evidence for a flare mechanism involving the reconnection of the field lines connecting the star and the disc or at the boundary between the stellar magnetosphere and the inner disc. 

One alternative explanation proposed to explain this lack of observational evidence is that star-disc reconnection events are frequent but weak. \citet{getman2021b} suggests that weaker X-rays may be hidden in the characteristic emission of the protostar. The flares they observed have lowest temperatures of the order of a few MK so our fiducial flare (1MK) can be considered low compared to this sample. However, we have seen that even if its weak it has a substantial effect on ionisation. A time-dependent model sampling a flare distribution is really necessary to provide quantitative estimates of the impact of these particles. The present work has to be understood as a proof-of-concept. 

\subsection{Towards a more predictive modelling}
In this work, we have shown that the ionisation rates produced by energetic particles from star-disc magnetic field interactions are several orders of magnitude higher than those usually calculated from X-rays or Galactic CRs. However, we again must emphasise that our results are based on a simplified model. The ionisation rates amplitude we have calculated are very localised around a particular magnetic field line. In reality several steps have to be considered to build a more quantitative model. First, the flare parameters (temperature, size of the reconnection region) have to be properly sampled. Second, a time-dependent model sampling this flare distribution has to be considered. Third, the flare location needs to be randomly sampled over some disc radii ranges. Fourth, a better particle propagation model accounting for mirroring-focusing effects and diffusion has to be considered. Fifth, hydrogen recombination has to be accounted between two successive flares. And sixth, the ionisation rate has then to be time-averaged over timescales corresponding to typical observation duration. Ideally, our model then has to be applied to an upgraded version of \textsc{ProDiMo} including magnetic configurations (which is under development) to study the particle transport as in \citet{padovani2011effects}. Putting our result in ProDiMo, we could investigate their impact on observables, e.g. molecular ions. %

\section{Conclusions}

The ionisation rate in the accretion discs of T Tauri stars plays a critical role in the development of the magnetic instabilities necessary to explain the accretion of surrounding matter onto the central star. The level of ionisation required to trigger these instabilities are difficult to explain. In this paper, we propose that the ionisation results from magnetic reconnection episodes, arising from the interaction between stellar and disc magnetic fields, taking place at the surface of the inner disc. We have estimated the ionisation rates produced by protons, electrons and secondary electrons accelerated from these flares. We have calculated the ionisation rates for a column density range between $10^{19}~\rm cm^{-2}$ to $10^{25} ~\rm cm^{-2}$ accounting for different magnetic field configurations. We find that the ionisation rate produced by our fiducial flare model (temperature of 1MK, purely vertical magnetic field, energetic particle distribution as a power-law with an energy index of 3) is 6 orders of magnitude (or more) higher than the rates produced by previously proposed ionisation sources (stellar X-rays, GCRs, radionuclides). For our fiducial case, we found ionisation rates $\zeta=10^{-9} ~\rm  s^{-1}$ at column densities of $10^{25} ~\rm cm^{-2}$, while the ionisation rate at this depth is of the order of $10^{-17} ~\rm  s^{-1}$ due X-rays produced in a $1~\rm MK$ stellar flares and GCRs and $10^{-18} ~\rm  s^{-1}$ by radionuclides.

Although we are aware that our assumptions lead to an overestimation, we show that this process can be a dominant one among the ionisation processes in the inner disc of T Tauri stars. As there are several parameters in our model which are difficult to constrain experimentally or observationally, we have conducted a comparative analysis of these parameters. The aim of this analysis is to give a range of flare parameters so that the ionisation rates produced are dominant over other ionisation sources. We anticipate that it is the case for (1) a reconnection process accelerating particles following an injection flux with a power law $j \propto  E^{-\delta}$ for $\delta < 6$, (2) flares with temperatures above 1 MK, (3) particles propagating along the field line with a ratio of the toroidal component to the poloidal component $b_{\rm g}=B_{\phi}/ B_{\rm pol} < 1$. We have seen that in-situ-accelerated energetic particles ionisation rates are always larger than X-ray ionisation rates as the radius of the flare location increases. Hence, ionisation by in-situ-accelerated energetic particles during magnetic reconnection events may be another mechanism to locally produce enhanced ionisation rates in the inner disc region of T Tauri stars. A more quantitative estimation is beyond the scope of this first model and deserves a future work in which time and space effects will be included.

\section*{Acknowledgements}
We acknowledge financial support from ``Programme National de Physique Stellaire'' (PNPS) and from "Programme National des Hautes Energies" (PNHE) of CNRS/INSU, France. This work has been conducted under the INTERCOS project of CNRS/IN2P3. CS thanks LUPM for hosting him during his long term visit (CRCT and d\'el\'egations CNRS). Ch. Rab is grateful for support from the Max Planck Society and acknowledges funding by the Deutsche Forschungsgemeinschaft (DFG, German Research Foundation) - 325594231. The authors thank T.P. Downes, J. Ferreira, G. Lesur, J. Morin, D. Rodgers-Lee, A. Vidotto, S. Xu for fruitful discussions. 


\section{Data Availability}
The data underlying this article will be shared on reasonable request to the corresponding author.



\bibliographystyle{mnras}
\bibliography{biblio} 








\bsp	
\label{lastpage}
\end{document}